\documentclass[twocolumn,graphics,floatfix,a4paper,superscriptaddress,aps]{revtex4-1}
\usepackage{color}
\usepackage[utf8]{inputenc}
\usepackage{graphicx}
\usepackage{amsmath}
\usepackage{amssymb}
\usepackage{placeins}

\usepackage{hyperref}
\hypersetup{
  pdfnewwindow=true, colorlinks=true,
  linkcolor=blue, anchorcolor=blue,
  citecolor=blue, filecolor=blue,
  menucolor=blue, urlcolor=blue}

\newcommand{\comment}[1]{\textit{}}
\newcommand{\bit}{\begin{itemize} \setlength{\itemsep}{0ex} \setlength{\topsep}{0ex} } %%%%%%% for draft %%%%
\newcommand{\eit}{\end{itemize}}
\newcommand{\be}{\begin{equation}}
\newcommand{\ee}{\end{equation}}
\newcommand{\bea}{\begin{eqnarray}}
\newcommand{\eea}{\end{eqnarray}}
\newcommand{\ba}{\begin{align}}
\newcommand{\ea}{\end{align}}

\providecommand{\gs}{\ensuremath{g^*_B}}

\begin{document}
\title{A unified numerical approach to topological 
  semiconductor-superconductor heterostructures}
\author{Georg W. Winkler} \email{georg.winkler@microsoft.com}
\affiliation{Microsoft Quantum, Microsoft Station Q, University of California, Santa Barbara, California 93106-6105 USA} 
\author{Andrey E. Antipov} \affiliation{Microsoft Quantum, Microsoft Station Q, University of California, Santa Barbara, California 93106-6105 USA}
\author{Bernard van Heck} \affiliation{Microsoft Quantum, Microsoft Station Q, University of California, Santa Barbara, California 93106-6105 USA} \author{Alexey
  A. Soluyanov} \affiliation{Physik-Institut, Universit\"at Z\"urich,
  Winterthurerstrasse 190, CH-8057 Zurich, Switzerland} \affiliation{Department of
  Physics, St. Petersburg State University, St. Petersburg, 199034,
  Russia} \author{Leonid
  I.  Glazman} \affiliation{Departments of Physics and Applied
  Physics, Yale University, New Haven, CT 06520, USA}\author{Michael
  Wimmer} \affiliation{QuTech and Kavli Institute of Nanoscience,
  Delft University of Technology, 2600 GA Delft, The Netherlands}
\author{Roman M. Lutchyn} \affiliation{Microsoft Quantum, Microsoft Station Q, University of California, Santa Barbara, California 93106-6105 USA} \date{\today}

\begin{abstract}
We develop a unified numerical approach for modeling semiconductor-superconductor heterostructures. 
% Our approach takes into account on equal footing important key ingredients: proximity-induced superconductivity, orbital and Zeeman effect of an applied magnetic field, spin-orbit coupling as well as the electrostatic environment in a realistic geometry. 
All the key physical ingredients of these systems --- orbital effect of magnetic field, superconducting proximity effect and electrostatic environment --- are taken into account on equal footing in a realistic device geometry. 
As a model system, we consider indium arsenide (InAs) nanowires with epitaxial aluminum (Al) shell, which is one of the most promising platforms for Majorana zero modes. We demonstrate qualitative and quantitative agreement of the obtained results with the existing experimental data. Finally, we characterize the topological superconducting phase emerging in a finite magnetic field and calculate the corresponding topological phase diagram.    

\end{abstract}

\maketitle

\section{Introduction}

The simulation of realistic quantum devices represents a challenging problem in computational physics, due to the difficulty of reaching quantitative agreement with experimental data on the basis of effective empirical models.
The complexity of the problem is particularly demanding in the case of hybrid superconductor-semiconductor devices, which are promising platforms for realizing Majorana zero modes~\cite{Kitaev2001, Moore1991, Nayak1996, Read2000, Ivanov2001,Sau2010, Alicea10, Lutchyn2010, Oreg2010, Nayak08, Alicea2012, Beenakker2013, Flensberg_review, DasSarma2015,Lutchyn17,Aguado17}, since semiconductors and superconductors ({\it i.e.}~metals) are described by disparate characteristic length scales, effective masses, and electron densities.
Furthermore, recent material science improvements of the semiconductor-superconductor interface~\cite{Krogstrup15, Lutchyn17}, due to the in-situ deposition of Al on InAs nanowire facets, reached the strong tunneling regime between the two materials~\cite{Chang2015, Deng2016, Nichele2017, Zhang2017, Sole2017}.
In this regime~\cite{Stanescu2011}, the wave functions of superconducting and semiconducting states are strongly hybridized due to the large coupling at the materials' interface.
The physical parameters required to describe the semiconducting material in simple effective models ~\cite{Lutchyn2010,Oreg2010} -- e.g.,~proximity-induced pairing, $g$-factor and spin-orbit coupling strength -- are strongly renormalized by such a hybridization~\cite{Cole2015, Sticlet2017, Antipov2018, Mikkelsen2018, Reeg2018}.
Thus, in order to understand physical properties of such a hybrid system, one has to develop a comprehensive numerical approach which takes into account different competing physical effects on equal footing.
The development of appropriate device simulations carrying out this task is crucial for topological quantum computing proposals~\cite{DasSarma2015, Lutchyn17} which rely on good control of Majorana zero modes. 

In this paper, we develop a unified numerical approach for modeling of the semiconductor-superconductor heterostructures.
Our approach appropriately takes into account the proximity-induced superconductivity in the strong tunneling limit, the orbital and Zeeman effect of an applied magnetic field, and the spin-orbit coupling, all within a self-consistent treatment of the electrostatic environment in the Thomas-Fermi approximation and in  a realistic geometry.
The simultaneous treatment of these ingredients was not achieved in previous numerical works to date~\cite{Stanescu2011,Prada2012,Rainis2013,Cole2015,Bas2016,Reeg2017,Sticlet2017,Reeg2018,Dmytruk2018}, which relied on effective phenomenological models neglecting the treatment of electrostatics and/or assuming the weak tunneling limit at the semiconductor-superconductor interface.
Several recent works developed Schrödinger-Poisson calculations for proximitized nanowires~\cite{Vuik2016, Antipov2018, Mikkelsen2018, woods2018,Dominguez2017,Escribano2018}, crucial for understanding electrostatics and gating effects, but they similarly neglected some of the other key ingredients listed above (most notably, the orbital effect of the magnetic field).
The present approach builds on the recently improved treatment of electrostatic effects in the strong tunneling limit~\cite{Antipov2018, Mikkelsen2018}, incorporating the orbital effect of the magnetic field~\cite{Bas2016, Winkler2017} as well as the dependence of the spin-orbit coupling on the external electric field.
We demonstrate that such a unified treatment is crucial for a better understanding of the large body of existing experimental data~\cite{Mourik2012,
  Rokhinson2012, Deng2012,Churchill2013, Das2012, Finck2012,
  Albrecht2016, Zhang2016, Chen2016, Deng2016, Suominen2017,
  Nichele2017, Gazibegovic2017, Zhang2017, Zhang2017a, Sestoft2017,Deng2017,Sole2017,Laroche2017,Chang2015,Michiel2018,Jie2018}. 

Tunneling conductance experiments~\cite{Chang2015,Deng2016,Nichele2017,Sole2017,Michiel2018} on InAs/Al~\cite{Krogstrup15} and InSb/Al~\cite{Gazibegovic2017} hybrid nanowires show that the high-quality epitaxial semiconductor-superconductor interface translates into a proximity-induced gap comparable to that of bulk Al, with very small sub-gap conductance at zero magnetic field.
These findings are evidence of the strong coupling between the two materials~\cite{Stanescu2011}.
At the same time, in a finite magnetic field, the sub-gap density of states appears to be strongly dependent on the gate voltage applied to the semiconductor~\cite{Sole2017,Michiel2018} (see Fig.~\ref{fig:fig1} for a representative gate geometry).
This dependence can be qualitatively attributed to the effect of the gate voltage on the electron density profile in the semiconductor. 
When the gate voltage is such that electrons are attracted away from the Al facets, the coupling to the superconductor is decreased and thus the low-energy states are more sensitive to the magnetic field.
In order to quantitatively capture this behavior, and following previous experimental works~\cite{Sole2017,Michiel2018,Albrecht2016}, it is convenient to define an ``effective $g$-factor'', $\gs=(2/\mu_B)\,(\partial E/\partial B)|_{B=B^*}$~\footnote{Note that the definition of the effective ``$g$-factor'' is different from the conventional definition since the slope is measured at finite magnetic field as opposed to infinitesimal magnetic field as required for defining a conventional $g$-factor. If dependence of $g(B)$ is linear, the two quantities coincide.}, which tracks the sensitivity of sub-gap states on external magnetic field.
Here, $B^*$ is the magnetic field at which we follow the evolution of the sub-gap states.
This ``effective $g$-factor'' can be close to that of Al ($|g_{\rm Al}|\approx 2$) if the wave function of the sub-gap states has large weight in the superconductor, and vice versa close to that of, say, InAs ($|g_{\rm InAs}|\approx 15$~\cite{Winkler2003}) if the wave function has large weight in the semiconductor.
Furthermore, $\gs$ can also be enhanced by the orbital effect of the magnetic field~\cite{Winkler2017}.
Understanding the renormalization of the $g$-factor is very important for studies of the topological phase diagram, since it determines the critical magnetic field for the topological phase transition.

Another important effect which motivated this study is the dependence of the spin-orbit coupling on the external electric field.
In the present work we calculate self-consistently the strength of the Rashba spin-orbit coupling as a function of an external gate voltage. The magnitude of Rashba spin-orbit coupling is difficult to extract in semiconductor-superconductor heterostructure.
In semiconductor nanowires, its value is typically inferred from the weak-anti-localization measurements~\cite{Weperen15,Shabani16}.
Here, instead, we use independent standard sources for the material parameters and calculate the dependence of spin-orbit coupling on the external electric field.

Overall, the approach developed in this work takes into account the important effects for understanding physical properties of the semiconductor-superconductor heterostructures: electrostatic environment, proximity-induced superconductivity, and orbital effect of an applied magnetic field.
It combines a number of previous approaches focusing on electrostatic effects~\cite{Vuik2016, Antipov2018, Mikkelsen2018, woods2018, Michiel2018} and treats the superconductivity explicitly rather than integrating out the corresponding degrees of freedom~\cite{Antipov2018,Mikkelsen2018,Bas2016,Reeg2017,Reeg2018, Wojcik2018, Michiel2018}.
Treating these two effects simultaneously with orbital effect~\cite{Bas2016, Winkler2017} allows one to understand the physical properties of semiconductor-superconductor heterostructures in a wide range of parameters.
Our results are important for interpreting the existing transport measurements as well as designing more complicated Majorana-based qubit experiments. 

The paper is organized as follows: We first introduce the setup, the
methodology and the model in Sec.~\ref{sec:methods}.
Then we present the obtained results.
In Sec.~\ref{sec:B0} we start by exploring the induced gap in the semiconductor in the absence of magnetic field ($B=0$).
In Sec.~\ref{sec:B1} we show results for finite magnetic field, namely the investigation of $\gs$ and of the topological phase diagram.
For concreteness, throughout the paper we use parameters for the InAs/Al heterostructure, although our approach is identically applicable also to other materials such as InSb/Al.

\section{\label{sec:methods}Setup and Methodology}

The simulations are separated in two parts.
In the first part, the electrostatic potential is computed by solving the Poisson equation for the wire cross-section (see Fig.~\ref{fig:fig1}).
In the second part, the potential is plugged into a realistic Bogoliubov-de Gennes Hamiltonian, whose eigenvalues and eigenstates are calculated as a function of the momentum along the wire as well as external parameters, like the applied magnetic field.
We assume that the electrostatic potential is independent of the applied magnetic field, and is thus calculated only at zero magnetic field. While due to the interplay of Coulomb interactions and Zeeman splitting the chemical potential in principle has a response to the magnetic field, in our case this effect will be suppressed due to the screening by the superconductor and comparatively large electron density in InAs~\cite{Vuik2016}.

\subsection{\label{electrostatics}Electrostatics}

The electrostatics of the system is determined by the
Poisson equation
\begin{equation}
  \nabla\cdot\left( \epsilon_r(\mathbf r) \nabla \phi(\mathbf r) \right) =
  \frac{\rho_\mathrm{tot}[\phi(\mathbf r)]}{\epsilon_0},
\end{equation}
where the total charge density $\rho_\mathrm{tot}[\phi(\mathbf r)]$ is
a functional of the potential $\phi(\mathbf r)$ and $\epsilon_r$ the
relative dielectric constant. We include four contributions to
$\rho_\mathrm{tot}$
\begin{equation}
  \rho_\mathrm{tot} = \rho_\mathrm{e} + \rho_\mathrm{hh} +
  \rho_\mathrm{lh} + \rho_\mathrm{acc},
\end{equation}
where $\rho_\mathrm{e}$, $\rho_\mathrm{hh}$ and $\rho_\mathrm{lh}$ are
the mobile charges of the conduction, heavy hole and light hole bands
of InAs. The charge density $\rho_\mathrm{acc}$ corresponds to fixed charges at the InAs surface (excluding the InAs-Al interface), e.g. due to surface chemistry of InAs. In our simulation $\rho_\mathrm{acc}$ is added to model the electron accumulation layer of the InAs surface~\cite{Olsson1996, Degtyarev2017} --- the details of which are discussed below. 

For mobile charges we use the Thomas-Fermi approximation for a 3D electron
gas~\cite{Kittel}
\begin{equation}
  \rho_\mathrm{e}(\phi) = -\frac{e}{3\pi^2}\left(\frac{2 m_\mathrm{e}
      \phi \theta(\phi)}{\hbar^2}\right)^{3/2},
  \label{eq:tfe}
\end{equation}
where $\theta$ stands for the Heaviside step function, corresponding to the Fermi-Dirac distribution at zero temperature. The
analogous expression for the holes is
\begin{equation}
  \rho_i(\phi) = \frac{e}{3\pi^2}\left(\frac{2 m_i (-E_0-\phi)
      \theta(-E_0-\phi)}{\hbar^2}\right)^{3/2},
  \label{eq:tfh}
\end{equation}
with $E_0$ being the band gap of the semiconductor and the index $i$
corresponds to the heavy hole (hh) and light hole (lh) band
respectively. The effective mass parameters and band gap of the semiconductor are taken from
Ref.~\onlinecite{Winkler2003}. The Thomas-Fermi approximation gives
potentials in very good agreement with a full Schr\"odinger-Poisson
treatment, as has recently been demonstrated in similar simulations of
InAs/Al heterostructures~\cite{Mikkelsen2018}.

\begin{figure}
  \includegraphics[width = \linewidth]{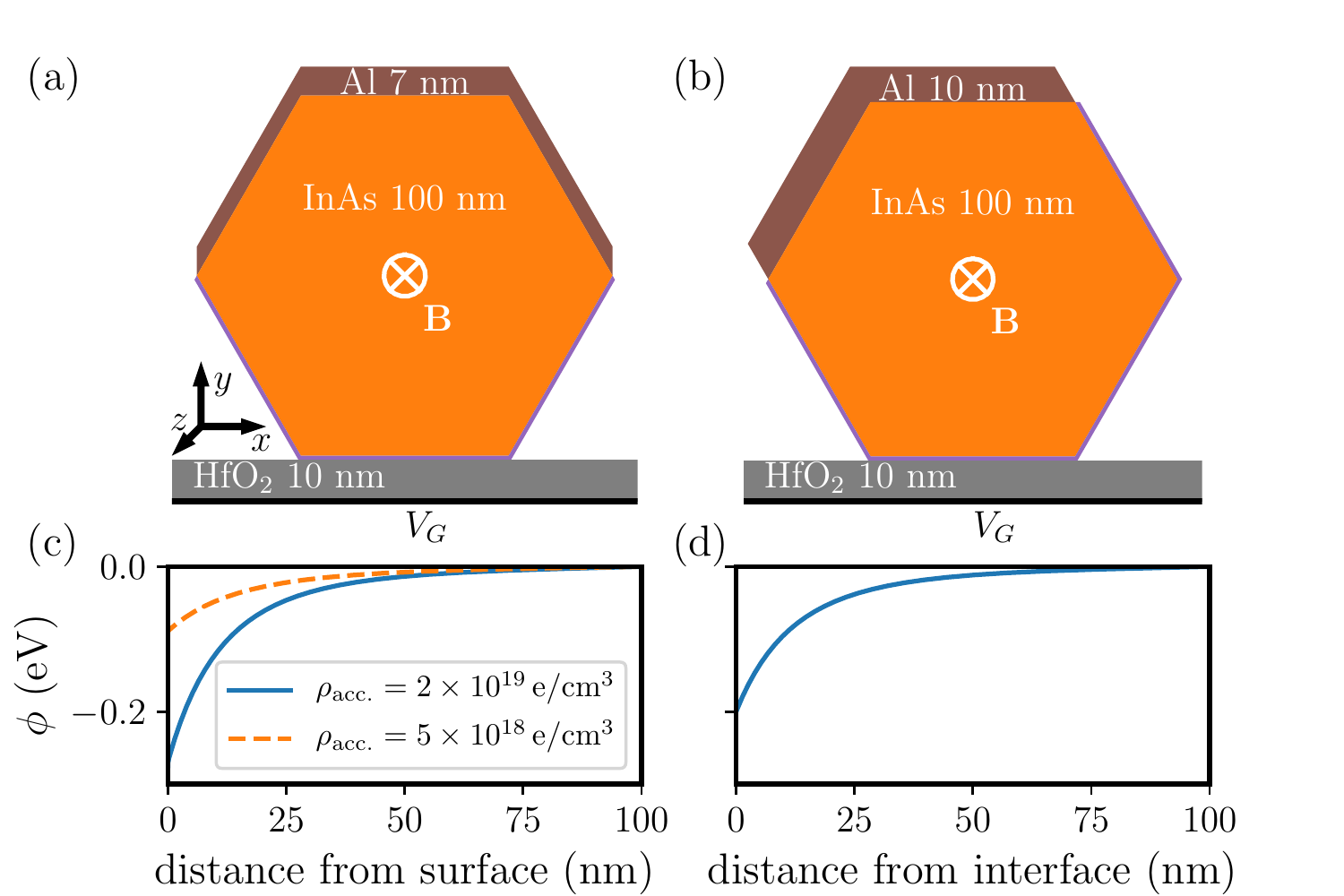}
  \caption{\label{fig:electrostatics} (a)  Electrostatic
    setup of the cross-section of a three-facet nanowire
    device: InAs (orange), Al (brown), positive charge $\rho_\mathrm{acc}$ (purple) and
    the dielectric HfO$_2$ (gray), below which the backgate is
    located. (b) Same as (a), but for a two-facet device.
    (c) Conduction band profile in a 1D simulation (with translation-invariance parallel to the interface) of InAs
    terminated by a 1\,nm layer of positive charge density on the left
    side. (d) Conduction band profile of InAs terminated by Al on the
    left side.\label{fig:fig1}}
\end{figure}

In Figs.~\ref{fig:electrostatics}~(a) and (b) we show the hexagonal cross sections of two nanowire devices covered by three and two facets of Al respectively, representing the devices experimentally investigated in Ref.~\cite{Sole2017}.
InAs is known to have a strong surface accumulation layer at the pristine InAs surface~\cite{Olsson1996,Degtyarev2017}.
We model [0001] wurtzite InAs nanowires, for which the precise parameters of the surface accumulation layer are presently
unknown.
Therefore, we choose the parameters compatible with existing measurements of different surfaces.
We model the accumulation layer by depositing a 1~nm layer of positive charge density $\rho_\mathrm{acc}$ on the InAs surface.
We simulate results for two cases: $\rho_\mathrm{acc}=2\times 10^{19}$\,$e$\,cm$^{-3}$~and~$\rho_\mathrm{acc}=5\times 10^{18}\,e$\,cm$^{-3}$.

For~$\rho_\mathrm{acc}=2\times 10^{19}$\,$e$\,cm$^{-3}$, the average charge density is about $10^{18}$\,$e$\,cm$^{-3}$
in the semiconductor wire at zero gate voltage, consistent with the reported electrical characterizations of such
wires~\cite{Thelander2010}.
In the absence of other charges and gates, the conduction band of InAs is pinned about 0.25\,eV below the Fermi
level at the surface [see the 1D simulation in Fig.~\ref{fig:electrostatics}~(c)].
To investigate the influence of the accumulation layer we simulate also $\rho_\mathrm{acc}=5\times 10^{18}\,e$\,cm$^{-3}$, for which the conduction band of InAs is pinned about 0.1\,eV below the Fermi level at the surface consistent with different characterizations of InAs nanowires~\cite{Degtyarev2017}.

In the case of the InAs/Al interface, we assume a similar conduction band offset to the Fermi level 
of 0.2\,eV, presumably resulting from the work function difference between the two materials.
The presence of band bending at the semiconductor-superconductor interface has recently been identified to be a crucial ingredient for the strong proximity effect~\cite{Antipov2018, Mikkelsen2018}.
This specific value of the conduction band offset is motivated by recent angle resolved photo-emission spectroscopy (ARPES) measurements of InAs with a thin layer of epitaxial Al~\cite{ARPES}.
In the electrostatic simulation the Al layer is assumed to be grounded.
It enters the Poisson equation only in the form of a Dirichlet boundary condition, which is set to 0.2\,V due to the work function difference to InAs [see also Fig.~\ref{fig:electrostatics}~(d)].
A backgate is located below the dielectric layer, set to the backgate voltage $V_G$.
On the remaining three boundaries of the system we assume Neumann conditions. For the dielectric constant of InAs and HfO$_2$ we
use the values 15.15 and 25 correspondingly~\cite{Wilk2001}.

\begin{figure}
  \includegraphics[width = \linewidth]{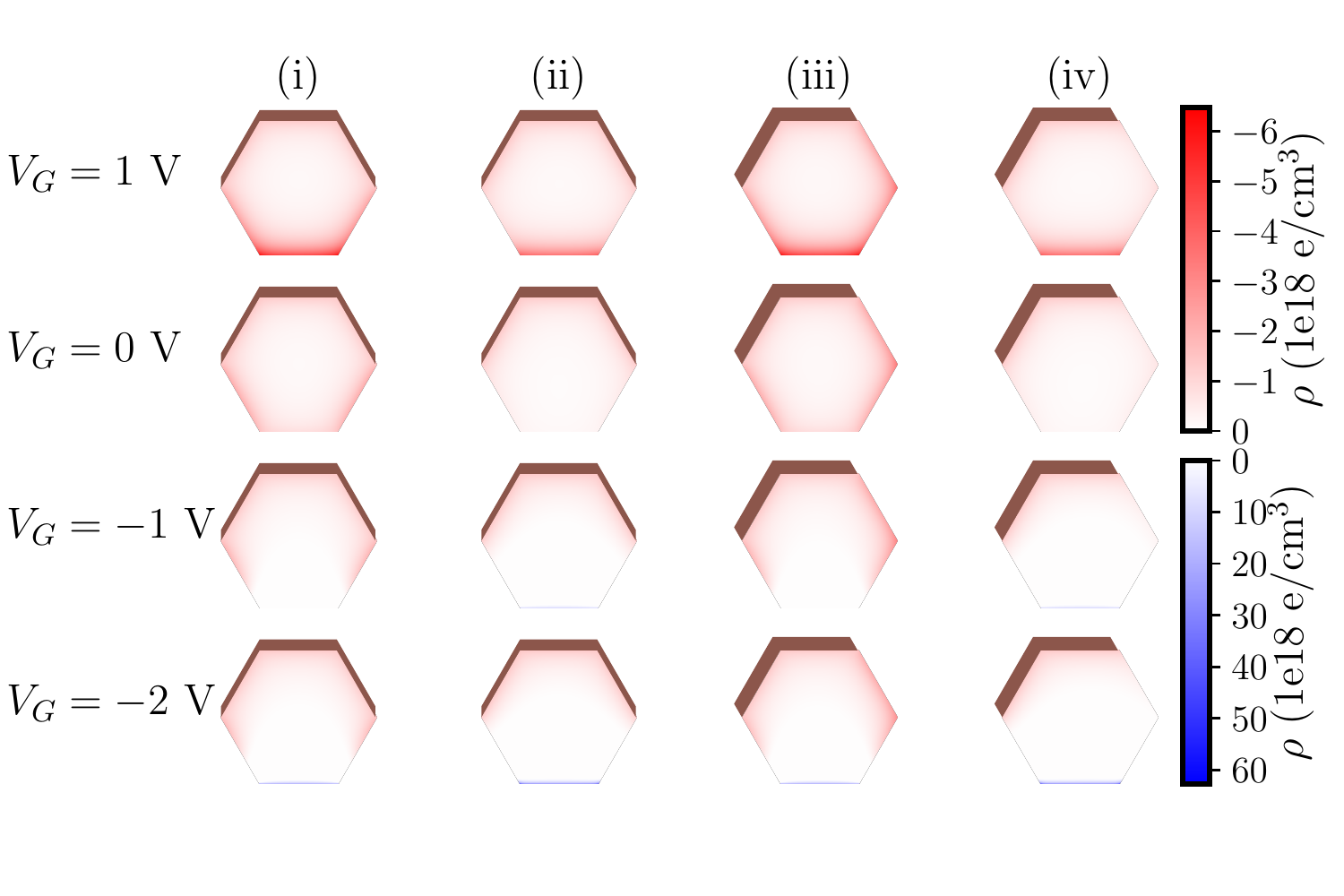}
  \caption{\label{fig:densities} Electron (red) and hole (blue) charge
    densities obtained from Eq.~\eqref{eq:tfe} and \eqref{eq:tfh} for
    different backgate voltages. (i) Three-facet device with
    $\rho_\mathrm{acc}=2\times 10^{19}$\,e/cm$^3$, (ii)
    three-facet device with
    $\rho_\mathrm{acc}=5\times 10^{18}$\,e/cm$^3$, (iii)
    two-facet device with
    $\rho_\mathrm{acc}=2\times10^{19}$\,e/cm$^3$, and (iv)
    two-facet device with
    $\rho_\mathrm{acc}=5\times 10^{18}$\,e/cm$^3$.}
\end{figure}

The resulting electron and hole densities are shown in
Fig.~\ref{fig:densities}, calculated from the self-consistent
electrostatic potential and Eq.~\eqref{eq:tfe} and \eqref{eq:tfh}. We
consider four different electrostatic configurations (i)-(iv),
corresponding to the two different wire designs with three and two
facets covered by Al and two different accumulation layers shown in
Fig.~\ref{fig:electrostatics}~(c).

Based on the density plots shown in Fig.~\ref{fig:densities} we
distinguish three generic cases for the localization of the electrons:
For $V_G>0$ we find most electrons to be localized close to the gate.
For $V_G\approx 0$ we see the electron density forming a ring of
charge at the surface of the semiconductor, resulting from the accumulation layer of InAs.  At more
negative voltages the only electrons left are near the InAs-Al interface.

Note that we find hole accumulation for $V_G<-1.5$\,V in configuration
(i, iii) [$V_G\leq-1V$ in configuration (ii, iv)] in
Fig.~\ref{fig:densities}.  This is a consequence of the small distance
of the backgate to the wire and of the small band gap of InAs. Due to
the large effective mass of the hole states they are very effective at
screening the gate once the potential becomes large enough for them to
be populated. While the screening length of the electrons is on the
order of 10-20\,nm, the screening length of the holes is about a
magnitude shorter on the scale of nm.  Going to more negative gate
voltage has little effect on the system since the gate is then
screened by the holes.

\subsection{Schr\"odinger solver}
After the electrostatic potential is calculated for a given geometry
and backgate voltage we plug it into the Schr\"odinger equation and
solve it for the cross-section of the device. The normal-state Hamiltonian for InAs/Al heterostructure reads
\begin{equation}
  \begin{aligned}
    H=&\left( \frac{\hbar^2}{2} \mathbf{k}^{T} m(\mathbf{r})^{-1}
      \mathbf{k}-E_\mathrm{F}(\mathbf{r})-e\phi(\mathbf{r}) \right)\sigma_0+\\
    &\frac{1}{2} \left(\boldsymbol{\alpha}(\mathbf{r}) \cdot \left( \boldsymbol{\sigma} \times
      \mathbf{k} \right) +\left( \boldsymbol{\sigma} \times
      \mathbf{k} \right) \cdot \boldsymbol{\alpha}(\mathbf{r}) \right)+ \\
      & Bg(\mathbf{r})\frac{\mu_B}{2}\sigma_z,
    \label{eq:ham0}
  \end{aligned}
\end{equation}
with the momentum-operator $\mathbf{k}^T = \left(-i\partial_x, -i\partial_y, k_z\right)$, the effective mass $m(\mathbf{r})$, the Fermi level $E_\mathrm{F}(\mathbf{r})$, the
electrostatic potential $\phi(\mathbf{r})$ (possibly including a disorder potential) and the
g-factor $g(\mathbf{r})$. Here $\sigma_i$ are Pauli matrices acting on spin degrees of freedom. We assume the wire to be aligned parallel to the $z$-direction and the Hamiltonian to be translation-invariant in $z$. The vector
$\boldsymbol{\alpha}(\mathbf{r})^T=\left( \alpha_x(\mathbf{r}), \alpha_y(\mathbf{r}), \alpha_z(\mathbf{r}) \right)$ contains the Rashba parameters $\alpha_x(\mathbf{r})$ and $\alpha_y(\mathbf{r})$ resulting from the electric field in the semiconductor, and the Dresselhaus parameter $\alpha_z(\mathbf{r})$ resulting from the bulk inversion asymmetry of wurtzite-InAs~\cite{Lew1996}.  The material parameters for InAs and Al are summarized in Tab.~\ref{tab:pars}. Note that when solving the Poisson equation we assume that Al is a perfect metal ({\it i.e.} impose the corresponding boundary conditions). Thus, $\phi$ is not solved for
inside of Al and is set to zero there. 

\begin{table}
  \caption{\label{tab:pars}%
    Material parameters for InAs and Al.  }
  \begin{ruledtabular}
    \begin{tabular}{ l c c}
      \textrm{Parameter}&
                          \textrm{InAs}&
                                         \textrm{Al}\\
      \colrule
      $m$ & 0.0229\,$m_0$~\cite{Winkler2003} & $m_0$ \\
      $\alpha_z$ & 0.03\,eVnm~\cite{Gmitra2016} & 0\,eVnm \\
      $g$ & -14.9~\cite{Winkler2003} & 2.0 \\
      $\Delta$ & 0\,meV & 0.34\,meV~\cite{Cochran1958}\\
      $E_\mathrm{F}$ & 0\,eV & 10\,eV
    \end{tabular}
  \end{ruledtabular}
\end{table}

We take into account superconductivity at the mean-field level. The corresponding Bogoliubov-de-Gennes (BdG) Hamiltonian in Nambu space~\cite{Tinkham} reads 
\begin{align}\label{eq:ham}
H_\mathrm{BdG}(\mathbf{r}, k_z) =& \begin{pmatrix} H(\mathbf{r}, k_z) & -i\sigma_y\Delta(\mathbf{r})
      e^{i\theta(\mathbf{r})} \\
      i\sigma_y\Delta(\mathbf{r}) e^{-i\theta(\mathbf{r})} & -H(\mathbf{r}, -k_z)^* \end{pmatrix}.
\end{align}
Here the superconducting pairing $\Delta$ is non-zero only in the superconductor (the superconducting phase $\theta$ is also only defined there).

The orbital effect is added to Eq.~\eqref{eq:ham} via the minimal
substitution
\begin{equation}
  k_i \rightarrow k_i - \frac{e}{\hbar} A_i,
  \label{eq:ms}
\end{equation}
with $\mathbf{A}$ being the vector potential. The BdG Hamiltonian
Eq.~\eqref{eq:ham} is gauge-invariant under transformations of the
form~\cite{Tinkham}
\begin{equation}
  \begin{aligned}
    \mathbf{A} \rightarrow & \mathbf{A}+\nabla \chi,\\
    \theta \rightarrow & \theta - \frac{2e}{\hbar} \chi.
  \end{aligned}
\end{equation}
The supercurrent in the superconductor is given by
\begin{equation}
  \mathbf{J}_S = -2e n_S \left( \frac{\hbar}{m} \nabla \theta +\frac{2e}{m} \mathbf{A} \right),
  \label{eq:sc}
\end{equation}
where $n_S$ is the superconducting density. Since in our simulations we
always deal with very thin superconducting films, we neglect any
screening effect of the superconductor and assume a homogeneous magnetic
field. The vector potential $\mathbf{A}$ and phase $\theta$ are chosen
such that Eq.~\eqref{eq:sc} leads to a physically correct distribution
of $\mathbf{J}_S$, see App.~\ref{sec:gauge} for details.

\begin{figure}
  \includegraphics[width = \linewidth]{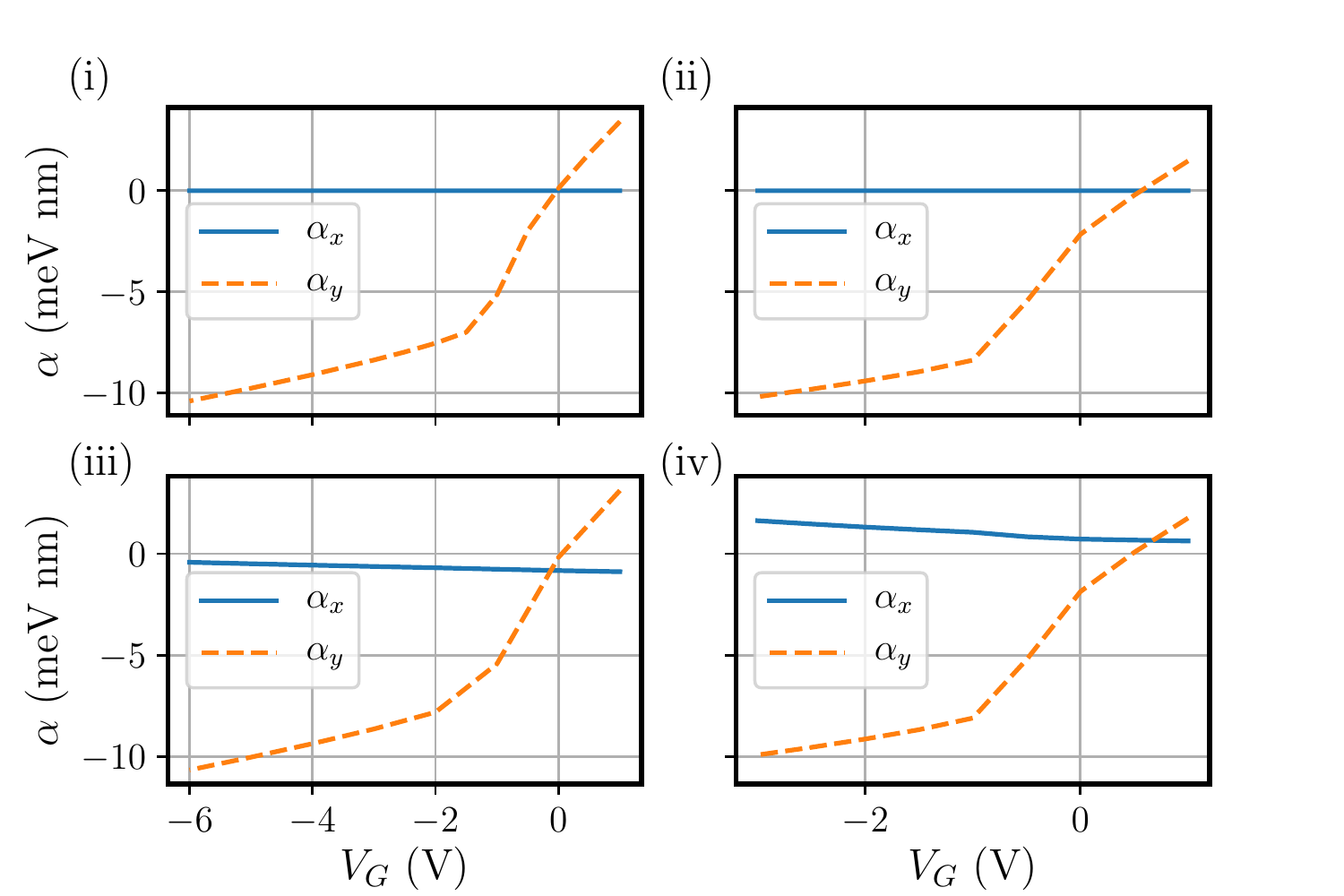}
  \caption{\label{fig:rashba} Bare Rashba parameters obtained from
    Eq.~\eqref{eq:rashba} as a function of gate voltage, corresponding
    to the four different electrostatic configurations in
    Fig.~\ref{fig:densities}~(i)-(iv)}
\end{figure}

The Rashba couplings $\alpha_x$ and $\alpha_y$ are nonzero only in the
semiconductor region and result from the symmetry breaking by the
electrostatic potential~\cite{Alexey2016}. We estimate the Rashba
couplings from the average electric field in the semiconductor region
from 8-band k.p-theory~\cite{Winkler2003}
\begin{equation}
  \alpha_i = \frac{eP^2}{3}\left[
    \frac{1}{E_0^2}-\frac{1}{{(E_0+\Delta_0)}^2} \right]
  \bar{\mathcal{E}}_i,
  \label{eq:rashba}
\end{equation}
where the average electric field in direction $i$ is obtained by
averaging $\mathcal{E}_i(\mathbf{r})$, calculated from the electrostatic potential $\phi(\mathbf{r})$, over the whole semiconductor region. This procedure neglects surface terms and the fact that wave functions might be localized more in the steep parts of the potential, therefore it gives a conservative, lower-bound, estimate for the Rashba spin-orbit coupling. The
following parameters for bulk InAs are used in accord with
Ref.~\onlinecite{Winkler2003}: $P=0.9197$\,eV\,nm, $E_0=0.418$\,eV and
$\Delta_0=0.380$\,eV. In Fig.~\ref{fig:rashba} we give the bare Rashba
parameters as a function of gate voltage for the different
devices. 

The Hamiltonian Eq.~\ref{eq:ham} is discretized on a quadratic mesh
 using the \texttt{kwant} package~\cite{kwant}. A lattice constant of 0.1\,nm is used for the discretization to
accommodate the small Fermi wavelength of Al. The interior eigenvalues close to the Fermi level of the sparse Hamiltonian are calculated with the shift-invert method as implemented in SciPy~\cite{scipy}, however, replacing the SuperLU algorithm packaged with SciPy with the more efficient MUMPS library~\cite{mumps1, mumps2}. With this numerical approach we are able to solve for interior eigenvalues of tight-binding Hamiltonians with several millions of basis states. 

\section{\label{sec:B0}Zero magnetic field}

A hard induced superconducting gap in the semiconductor is one of the
prime achievements of epitaxial InAs/Al hybrid
nanowires~\cite{Chang2015}. In these devices typically two to three
facets of the hexagonal wire are covered by a layer of Al. To achieve
critical magnetic fields that are large enough to form MZMs the
superconducting layer is kept very thin, between 5 to 10\,nm in
typical devices. While the superconducting layer is very thin it also has a very large density of states (DOS). Therefore,
in a hybrid device the induced gap in the semiconductor is often found to be of the same
order as the superconductor gap~\cite{Krogstrup15}.

\begin{figure}
  \includegraphics[width = \linewidth]{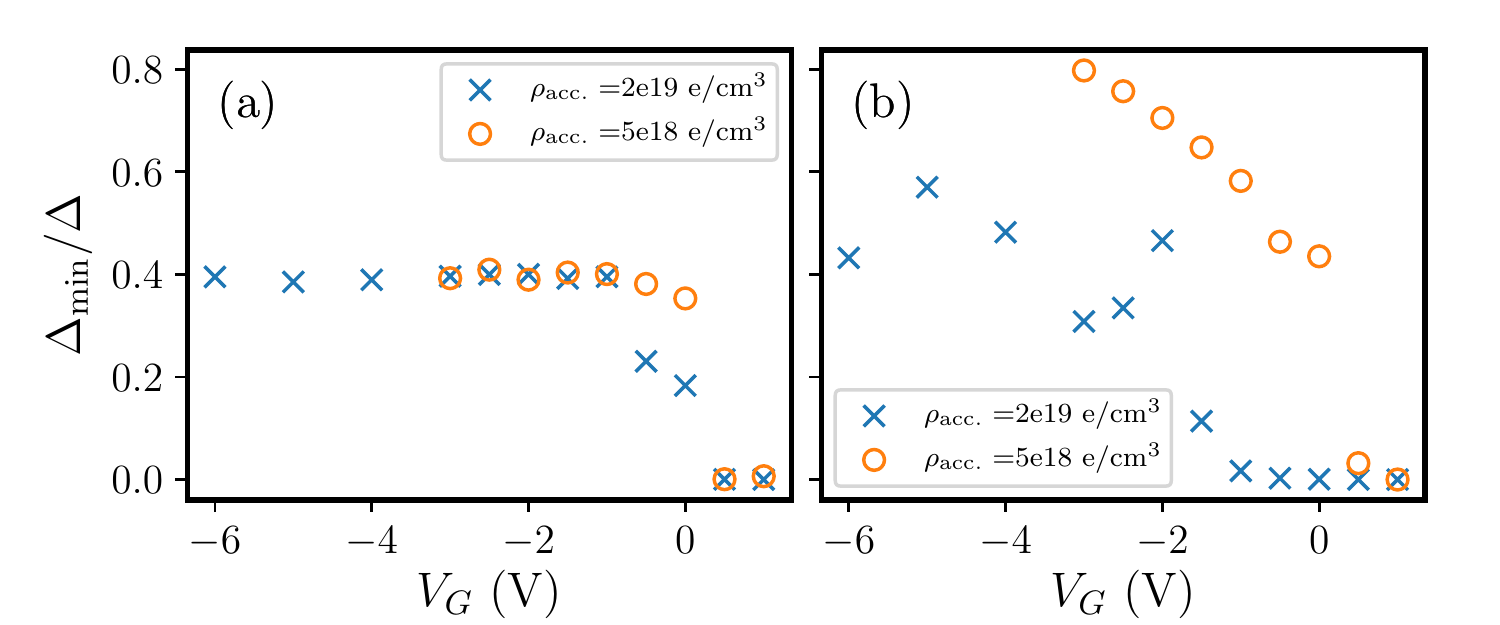}
  \caption{\label{fig:igall} Induced gap as a function of backgate
    voltage for the different devices and configurations. (a)
    Comparison of three-facet devices with 7 nm Al
    shell for the different accumulation layer strengths. (b) Same as (a), but for a two-facet device with a 10 nm Al shell.}
\end{figure}

The combined results for the minimal induced gap are shown in
Fig.~\ref{fig:igall}. All devices have large induced gap for
appropriate gate voltages. We now discuss the four different
electrostatic configurations presented in Fig.~\ref{fig:densities} and
the effects of disorder in detail.

\subsection{Three-facet device}

\begin{figure}
  \includegraphics[width = \linewidth]{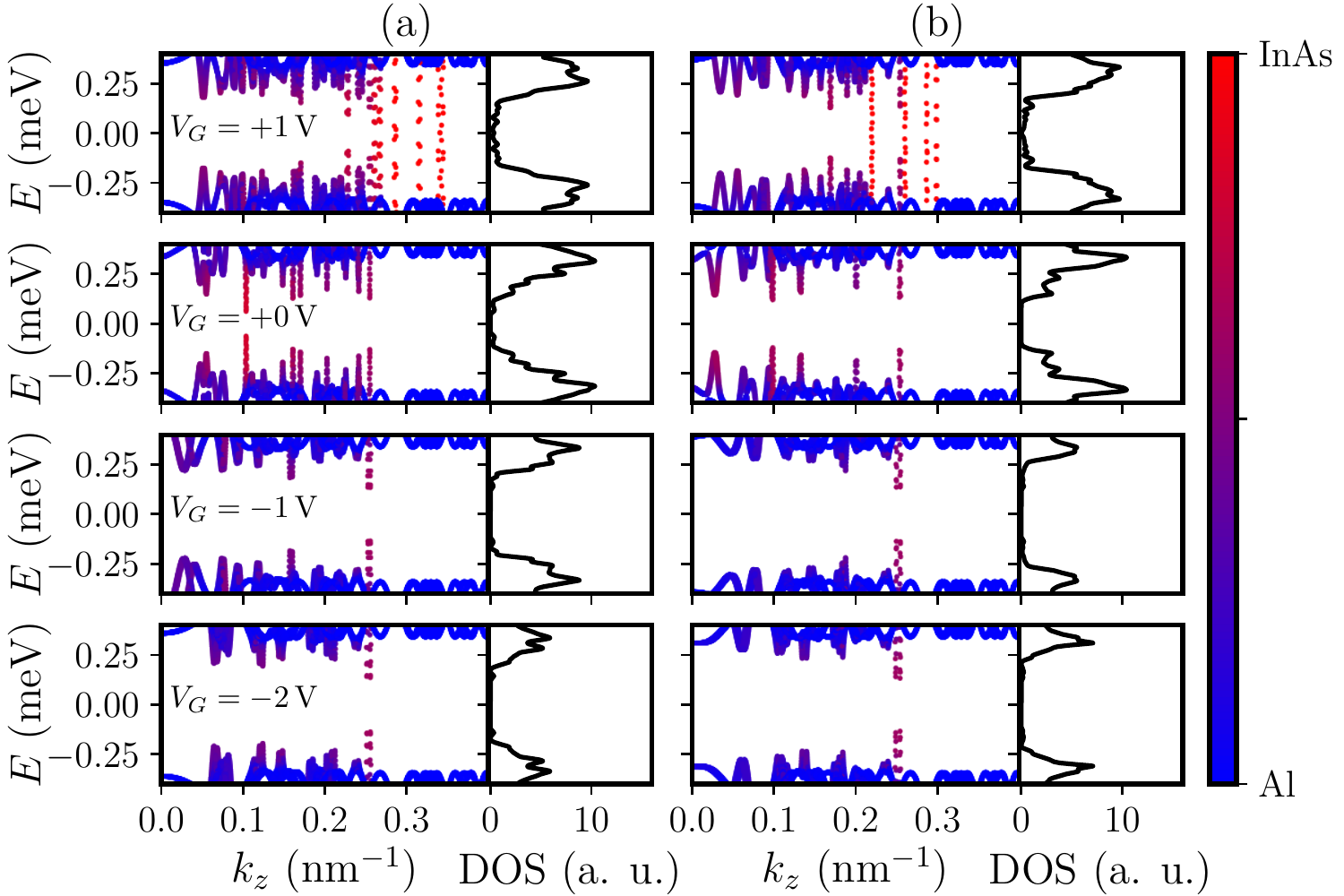}
  \caption{\label{fig:ig} (a) Induced gap in the three-facet
    nanowire with 7 nm Al shell and
    $\rho_\mathrm{acc}=2\times 10^{19}$\,e/cm$^3$, for different values of the backgate voltage.
    The one-dimensional energy spectrum and integrated one-dimensional DOS in the InAs region is shown. The color scale indicates the weight of the wave functions in the InAs and Al regions. For the dispersion and $k$ integration the point spacing is
    $dk=2\times10^{-4}$\,nm$^{-1}$. (b) Same as (a), but for $\rho_\mathrm{acc}=5\times 10^{18}$\,e/cm$^3$.}
\end{figure}

In Fig.~\ref{fig:ig} we show the energy spectrum and DOS in the InAs
region for different backgate voltages in the three-facet wire. For
all backgate voltages we find semiconductor-like states that have a strong
hybridization with the superconductor. These states appear below the Al gap, but
have a strong hybridization with Al as indicated by the color in
Fig.~\ref{fig:ig}. The DOS in the semiconductor is
obtained by integrating the band structure over momentum, multiplying
each eigenstate by its weight in the semiconductor~\footnote{We cut-off the integration at $k_\mathrm{max}=0.4\,\mathrm{nm}^{-1}$ since the weight in the semiconductor is negligible for larger momenta.}. A temperature broadening of
the energy levels of 50\,mK is assumed. The calculated DOS is consistent with experiment~\cite{Chang2015,
  Gazibegovic2017}: For negative gate voltages we typically find a hard gap, with the position of the coherence peaks showing little dependence on gate voltage. For positive gate voltages non-superconducting states enter the gap.

At positive backgate voltages, we find accumulation of electrons near
the backgate, on the opposite side of Al.  These states 
live almost 
completely in the InAs region and have negligible hybridization with the 
superconductor and thus no, or very small, induced gap. They contribute to a subgap 
conductance
for $V_G > 0$.

Around $V_G \approx 0$ the electron density is distributed along the surface of the semiconductor (see Fig.~\ref{fig:densities}). In this regime all
states have non-zero hybridization with the superconductor and a hard gap opens
up.

For sufficiently negative backgate voltage $V_G < 0$ the only states
left are in close proximity to the
superconductor. These are characterized by strong hybridization and induced gap on
the order of the superconductor gap. Note that a single state at
$k\approx 0.25$\,nm$^{-1}$ has significantly smaller hybridization and
induced gap than the other states in Fig.~\ref{fig:ig}~(a) and (b) for
$V_G \leq -1$\,V. Furthermore, note that our Schr\"odinger solver only
includes the electrons, hence no hole states show up in the DOS and
band structures of Fig.~\ref{fig:ig}. According to
Fig.~\ref{fig:densities} hole accumulation would be expected for
$V_G\leq-2$\,V for Fig.~\ref{fig:ig}~(a) [$V_G\leq-1$\,V for (b)].

\subsection{Effect of disorder on the induced gap}

Some previous attempts at simulating the superconducting proximity effect by treating the semiconductor and superconductor on equal footing often found an induced gap that is strongly dependent on geometric and microscopic details and 
significantly smaller than the one reported in experiments~\cite{Reeg2017, Mikkelsen2018}.
This is a consequence of the constraint imposed by momentum conservation at a smooth interface between semiconductor and superconductor. 
In such a case, tunneling between the two subsystems is suppressed due to energy and momentum constraints. Indeed, in this case tunneling rate, which is relevant for the proximity
effect~\cite{Stanescu2011}, is effectively proportional to one-dimensional DOS and decreases with 
$E_F$. 
Since $E_F$ is large in metals such as Aluminum, at any given parallel momentum the phase space which satisfies both constraints is small.
In other words, the level spacing coming from one-dimensional sub-band quantization in the superconductor is several orders of magnitude larger than the superconducting gap for Al films with a thickness of 10\,nm.
This results in a strong and non-monotonic dependence of the induced gap on the thickness of the superconductor~\cite{Mikkelsen2018}.
This dependence on the thickness of the superconductor is not observed in experiment and is an artifact of a parallel-momentum-conserving approximation at the surface. In fact, experimentally the opposite effect is observed, that the gap is enhanced for thinner Al thicknesses~\cite{Meservey1964,Meservey1971}.

Ref.~\cite{Antipov2018} demonstrated that disorder in the superconductor enhances the induced gap dramatically and, provided it is sufficiently strong, removes the non-monotonic dependence on the thickness of the superconducting layer.
Since a fully three dimensional simulation of a semiconductor-superconductor heterostructure would be extremely challenging, the disorder potential is chosen independent of $z$, such that the momentum $k_z$ parallel to the nanowire axis remains conserved.
This disorder breaks the momentum conservation in one direction parallel to the interface so that the semiconductor is now hybridized by an effectively two dimensional superconductor. The level spacing resulting from two-dimensional quantization in the superconducting film is smaller than the superconducting gap in the hybrid device geometries investigated here.
In reality, three-dimensional disorder in the superconductor breaks the momentum conservation also in the longitudinal direction, possibly enhancing the hybridization and the induced gap even further. Our two-dimensional disorder model is sufficient to qualitatively describe the effect of disorder on the hybridization of the semiconductor with the superconductor. However, in the topological phase a three-dimensional model of disorder will also lead to a number of new qualitative features such as impurity-induced subgap states~\cite{Lutchyn2012, Cole2016, Liu2018}, which is not captured by our model. We also do not consider here disorder in the semiconductor. High-quality semiconductor nanowires have been shown to be very clean with a mean free path of the order a micron~\cite{Lutchyn17}. Both of these effects are outside the scope of this work.

Furthermore, we note that in our simulations the momentum conservation at the semiconductor-superconductor interface is already broken due to the non-planar (i.e. hexagonal) shape of the Al shell.
We find that this has very similar effects as adding disorder to the system,~i.e. we find large induced gaps that depend only weakly on the geometry and thickness of the superconducting layer. 

While the semiconductor and also the interface to the superconductor is very clean~\cite{Krogstrup15}, the Al is naturally covered by an amorphous oxide
layer. This oxide layer provides a motivation to investigate --- additionally to the non-planar interface --- the effect of disorder on the surface of the Al shell. Motivated by this physical model, we restrict the disorder to a layer of 2\,nm thickness from
the outside of the shell. We use random on-site chemical potential in the specified region to simulate disorder. The disorder potential itself is  sampled from
a Gaussian distribution with a variance of 1\,eV. The disorder strength
and thickness are chosen such that an electron experiences on
average one scattering event while being reflected at the
Al-vacuum interface, corresponding to a mean-free path of  the order of the thickness of the superconductor.  

\begin{figure}
  \includegraphics[width = \linewidth]{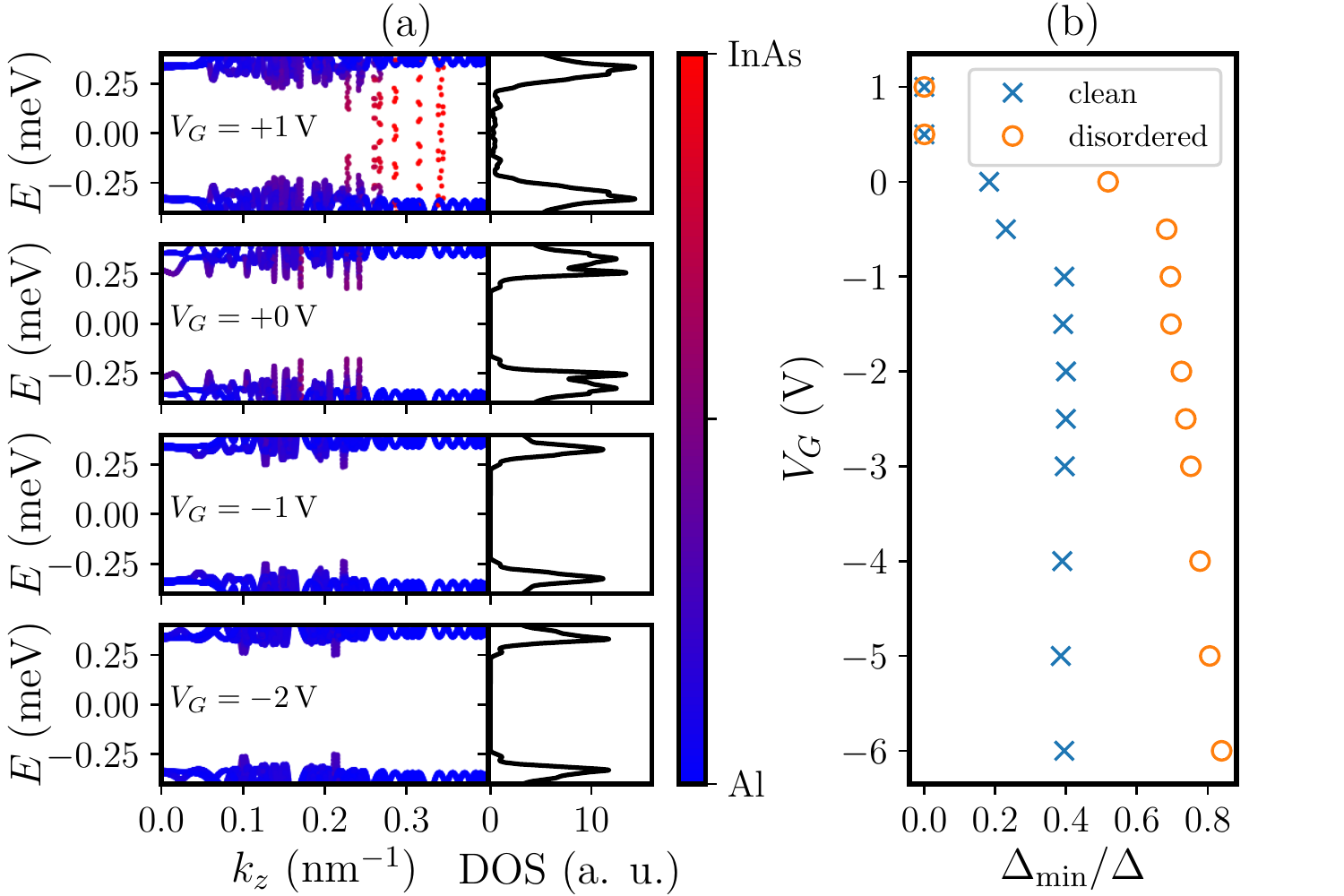}
  \caption{\label{fig:igdisorder} (a) Induced gap in the three-facet
    nanowire with 7 nm Al shell and
    $\rho_\mathrm{acc}=2\times 10^{19}$\,e/cm$^3$ with
    disorder. The one-dimensional energy spectrum and integrated
    one-dimensional DOS in the InAs region is shown. The weight of the
    states in the InAs and Al region is indicated by the color. For
    the dispersion and $k$ integration the point spacing is
    $dk=2\times10^{-4}$\,nm$^{-1}$. (b) Comparison of the minimal
    induced gap with and without disorder.}
\end{figure}

In Fig.~\ref{fig:igdisorder} we show that disorder is able to
significantly enhance the size of the induced gap. 
With disorder, all states close to Al show very strong
hybridization and induced gap on the order of 80\% of the Al gap at negative gate voltages.  On the other hand, the states contributing to the subgap conductance at positive gate voltages are not affected by the presence of disorder in Al, consistent with them having no weight in the superconductor.

\subsection{Two-facet device}

\begin{figure}
  \includegraphics[width = \linewidth]{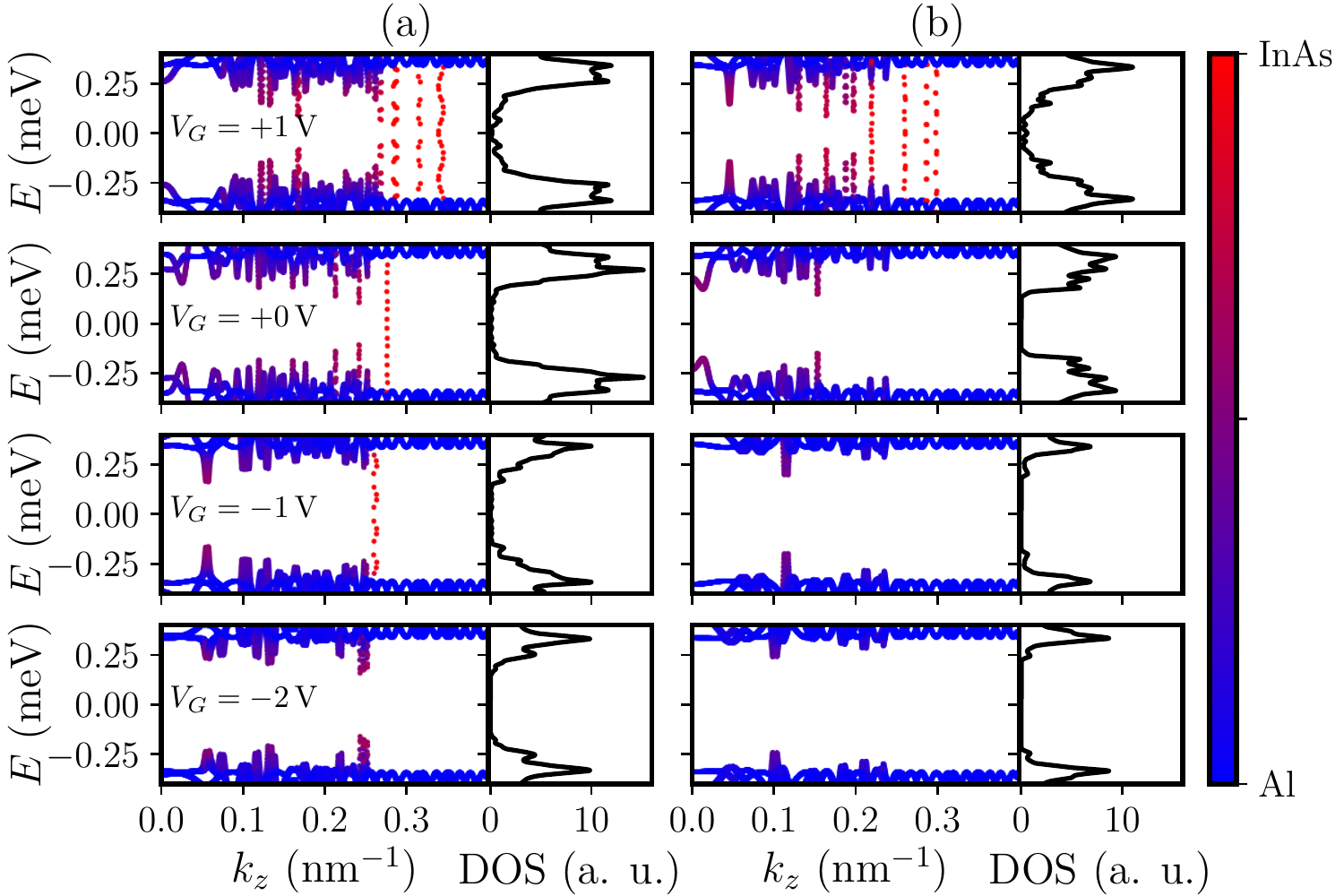}
  \caption{\label{fig:igtf} Induced gap in the two-facet nanowire
    depending on backgate voltage. The one-dimensional energy spectrum
    and integrated one-dimensional DOS in the InAs region is
    shown. The weight of the states in the InAs and Al region is
    indicated by the color. For the dispersion and $k$ integration the
    point spacing is $dk=2\times10^{-4}$\,nm$^{-1}$. The potentials
    correspond to Fig.~\ref{fig:densities}~(iii) (a) and (iv) (b).}
\end{figure}

In Fig.~\ref{fig:igtf} we show the electronic states and DOS in the
two-facet device for different backgate voltages. In general, we find
very similar results to the three-facet device.

We find that the backgate voltage at which the gap becomes hard is very
different for the different accumulation layers, with the threshold
being $V_G < -1.5$\,V for
$\rho_\mathrm{acc}=2\times 10^{19}$\,e/cm$^3$ and $V_G \leq 0$\,V
for $\rho_\mathrm{acc}=5\times 10^{18}$\,e/cm$^3$. The reason
for this becomes apparent when comparing the density distributions of
the two cases in Fig.~\ref{fig:densities}. Due to their accumulation
layer hexagonal InAs nanowires have a tendency to accumulate a higher
density in their corners than below their facets~\cite{Degtyarev2017}.
In the two-facet device there is a corner that is not adjacent to the
Al-shell or the backgate. The states living in this corner have 
no weight in the superconductor analog to the gate states and therefore 
result in a reduced gap compared to the three-facet device. 
In the strong accumulation layer case
$\rho_\mathrm{acc}=2\times 10^{19}$\,e/cm$^3$ large
 negative backgate voltage $V_G < -1$\,V is required 
 to open a superconducting gap for these states.

For the weak accumulation layer case
($\rho_\mathrm{acc}=5\times 10^{18}$\,e/cm$^{3}$) one finds even
enhanced induced gap compared to the three-facet device at large
negative gate voltage, although, comparable to the gap of the three-facet device with disorder. The induced gap in the clean two-facet device is
comparable to the three-facet device with disorder, see
Fig.~\ref{fig:igall}. This might be related to two effects: First, the two-facet device has less symmetry
than the three-facet device, as it lacks the vertical
mirror symmetry. Removing this symmetry allows a stronger
hybridization of semiconductor and superconductor states. Another difference is the thickness of the superconducting layers for the two
devices, which is investigated in App.~\ref{sec:thickness}. We find that a thicker superconductor also hybridizes the semiconductor more strongly.

\section{\label{sec:B1}Finite magnetic field}

Upon turning on a magnetic field parallel to the wire, and for appropriate gate voltages, the system can transition into the topological phase supporting MZMs~\cite{Lutchyn2010, Oreg2010}.
It is desirable for the critical field of the topological transition to be as small as possible, and in particular to be much lower than the critical field of the superconducting (e.g. Al) shell.
In this section, we first investigate the bulk DOS as a function of magnetic field and the behavior of $g^*_B$ as a function of gate voltage, which is important to determine the magnitude of the critical field of the topological transition.
Then, we investigate the topological phase diagram.

\subsection{Density of states}

\begin{figure}
  \includegraphics[width = \linewidth]{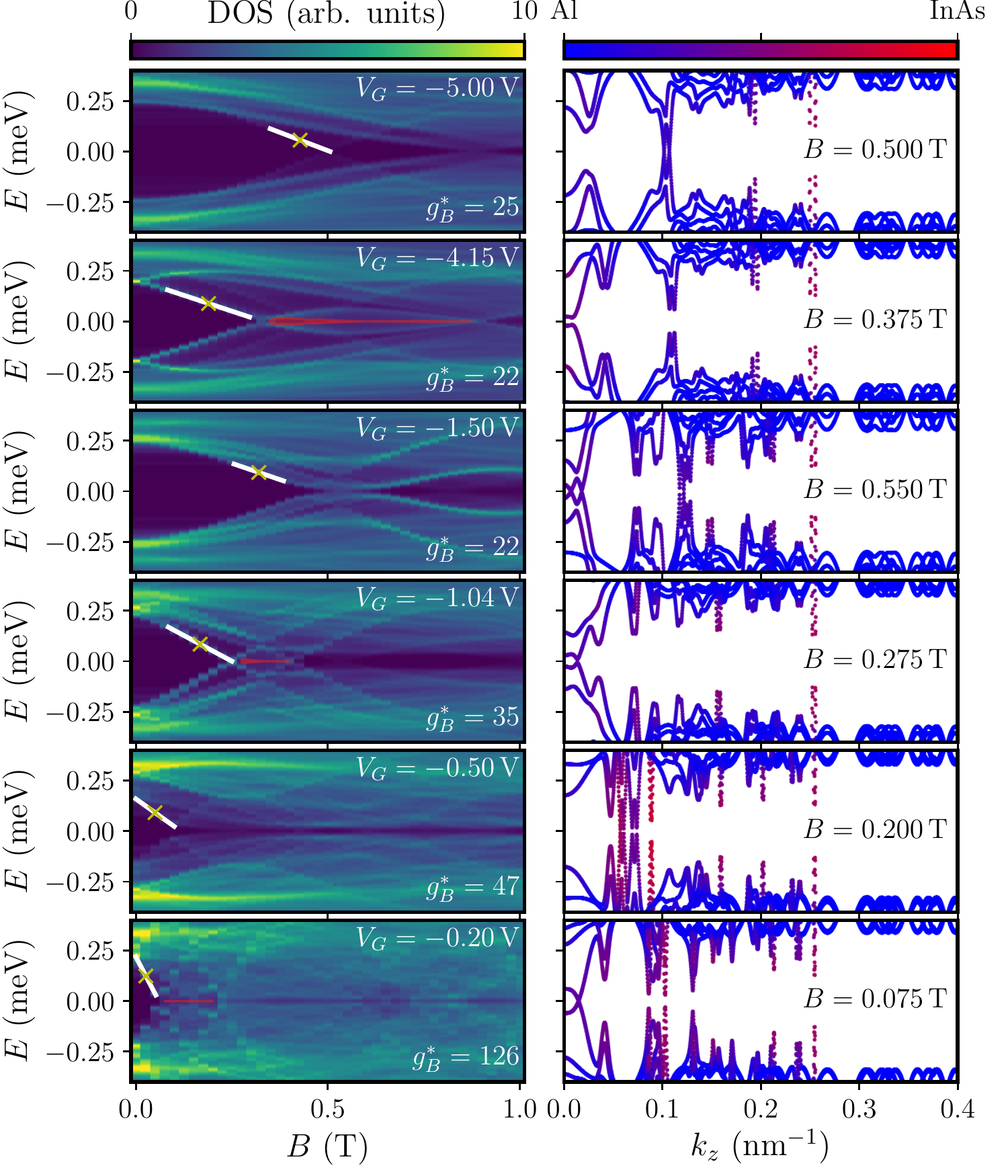}
  \caption{\label{fig:fb} DOS as a function of B for different
    backgate voltages in the three-facet device with 7 nm Al shell and
    $\rho_\mathrm{acc}=2\times 10^{19}$\,e/cm$^3$. The band
    structure is shown for a selected magnetic field strength. For
    $V_G = \{-4.15, -1.04, -0.20\}$\,V the device enters the
    topological phase for a range of magnetic fields. In this case the
    size of the topological gap is indicated in red.}
\end{figure}

Transport experiments in Majorana nanowires measuring the differential
conductance can be related to the local DOS at the tunnel contact if
the tunneling rate is small~\cite{Datta1997}. In our simulations we
calculate the DOS in the middle of an infinite wire. Therefore, no MZMs are visible
 with the only sign of the topological phase
transition being the bulk gap closing. Furthermore, 
we expect that the experimentally obtained differential
conductance will show bulk states with different relative intensity than in
 bulk DOS simulations~\cite{Datta1997}.

In Fig.~\ref{fig:fb} we show the bulk DOS as a function of magnetic
field and energy for six different backgate voltages. Both cases, where the system undergoes a topological phase transition and
where it stays trivial in the range of magnetic fields, are presented. In the
topological cases we indicate the topological gap in red. 
Due to the very large size of the Hamiltonian matrix it is impossible to
directly evaluate the Pfaffian to determine the topological
phase~\cite{Kitaev2001}. Instead, we use the fact that the gap of the Hamiltonian at
$k=0$ closes, always and only at topological phase
transitions. Considering further that the system must be in the
topologically trivial phase for $B=0$, one can determine the location of the topological phase this way. 

Regardless of topological character, we define similar to the experiment Ref.~\cite{Sole2017}
$\gs$ as the slope of the lowest-energy states represented in the DOS. For the cases in
Fig.~\ref{fig:fb}, in which the system undergoes a topological phase
transition, the gap closes linearly with $B$ and $\gs$ is
 independent of $B^*$ at which it is extracted. In the non-topological cases, however, the interplay of
spin-orbit coupling and finite chemical potential typically lead to a
nonlinear gap dependence with $B$~\footnote{In the non-topological cases the gap becomes extremely small, but does not close in general.}. In these cases we choose the magnetic field $B^*$
such that $\gs$ is maximized.

\begin{figure}
  \includegraphics[width = \linewidth]{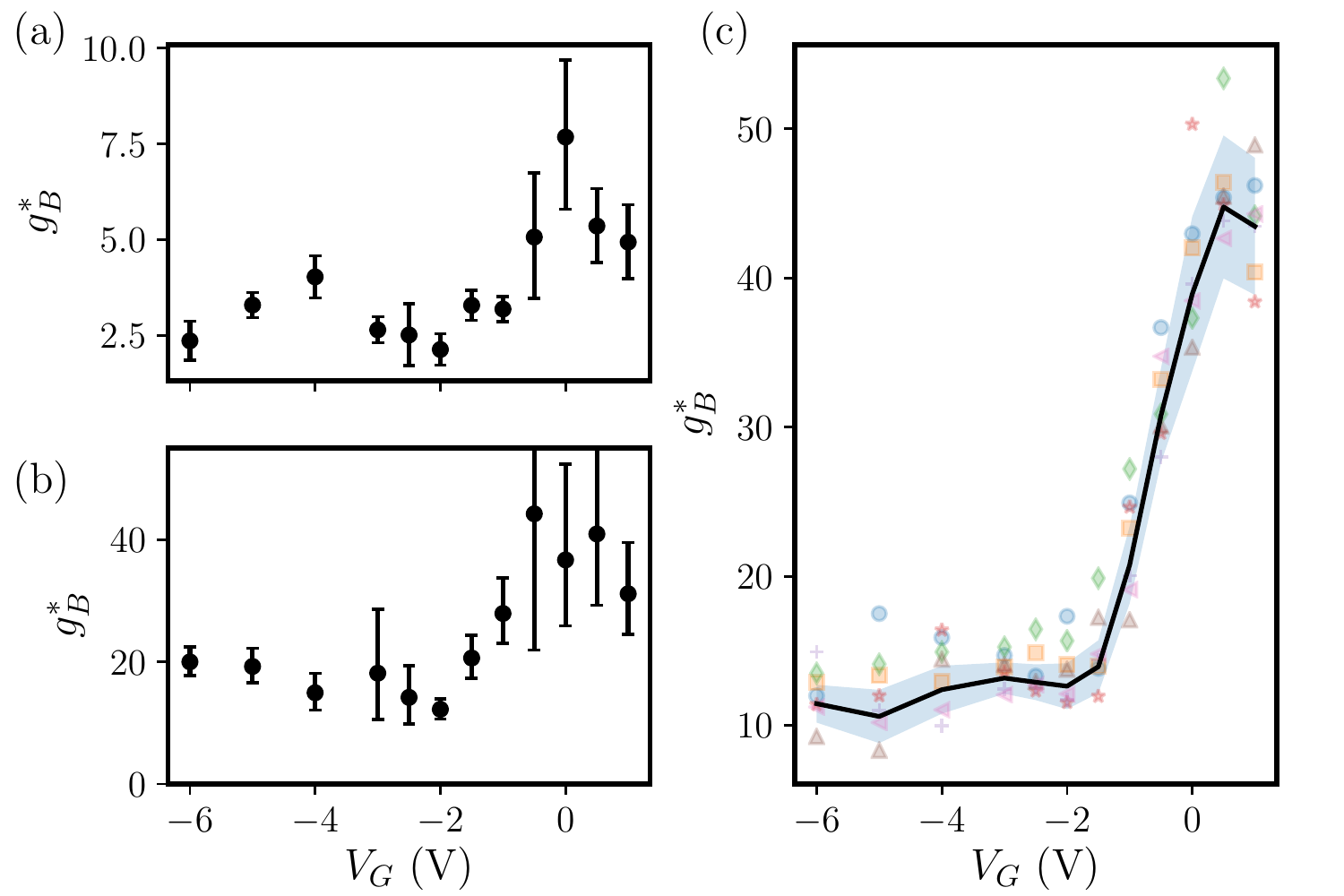}
  \caption{\label{fig:gfactor} Dependence of $\gs$ on the backgate
    voltage in the three-facet device with 7 nm Al shell and
    $\rho_\mathrm{acc}=2\times 10^{19}$\,e/cm$^3$ for (a)
    without orbital effects, (b) with orbital effects and (c) with
    orbital effects and disordered Al, averaged over seven
    disorder realizations which are shown faintly in the background, with identical parameters as in
    Fig.~\ref{fig:igdisorder}. The value of $\gs$ is obtained from
    fitting the magnetic field dependent bulk DOS closing. The error
    bars are obtained by assuming an uncertainty of $10\,\mu$eV in
    energy and 0.05\,T in magnetic field during the fit.}
\end{figure}

We observe a clear trend in the dependence of $\gs$ on the backgate voltage, see 
Fig.~\ref{fig:gfactor}. In Fig.~\ref{fig:gfactor}~(a) we show
results for $\gs$ without the orbital effect of magnetic field. Since
$\gs$ is directly affected by the renormalization of the $g$-factor,
it is an indicator of the coupling strength of semiconductor to superconductor. In general,
we find a non-monotonic behavior, resulting from the different
coupling strengths of the different subbands, although showing already
a trend in $\gs$, with large values of $\gs>5$ only at positive or small
negative backgate voltages. The trend becomes even more pronounced in
Fig.~\ref{fig:gfactor}~(b) where the orbital effect of magnetic field
is included. This is a result of the orbital effect being very
sensitive to the extent of the wavefunction, which is tuned by the
backgate. For positive, or small negative, backgate voltages the slope
$\gs$ is significantly larger than the bulk $g$-factor of InAs,
corresponding to an orbital enhancement of $g$-factor as discussed in
Ref.~\cite{Winkler2017}. At this gate voltages states are localized in the accumulation layer, along the surface of InAs. These states are strongly influenced by the orbital effect since
they are easily penetrated by magnetic flux. For large negative gate voltages, however, the wave functions
are confined close to the superconductor, and the orbital effect is
suppressed. 

In Fig.~\ref{fig:gfactor}~(c) we investigate the effect of disorder,
for the same system as in Fig.~\ref{fig:igdisorder}. We find that
while $\gs$ is more dependent on the disorder realization than the
induced gap, the typical variation is often still less than the error
bar of our fit. The main effect of disorder is that it further
enhances the coupling of semiconductor with superconductor, resulting in effective g-factors smaller than in the case without disorder. Also the step-like
behavior of $\gs$ with backgate voltage becomes more pronounced, being
very close to the one measured in experiments~\cite{Sole2017}.  Still,
even after taking disorder into account we find that our values of
$\gs$ are often overestimated compared to the experiment. This might
be a limitation of our Hamiltonian Eq.~\eqref{eq:ham}, which does not
include confinement effects on the InAs $g$-factor and the effective
mass~\cite{Stein1983, Lommer1985,Ivchenko1992}. Implementing a more
sophisticated Hamiltonian, like the 8-band Kane
Hamiltonian~\cite{Kane1957}, might be required for more accurate
results, although it would be very difficult due to the prohibitive computational cost resulting from the very dense discretization.

\subsection{\label{sec:interference}Magnetic interference effects}

\begin{figure}
  \includegraphics[width = \linewidth]{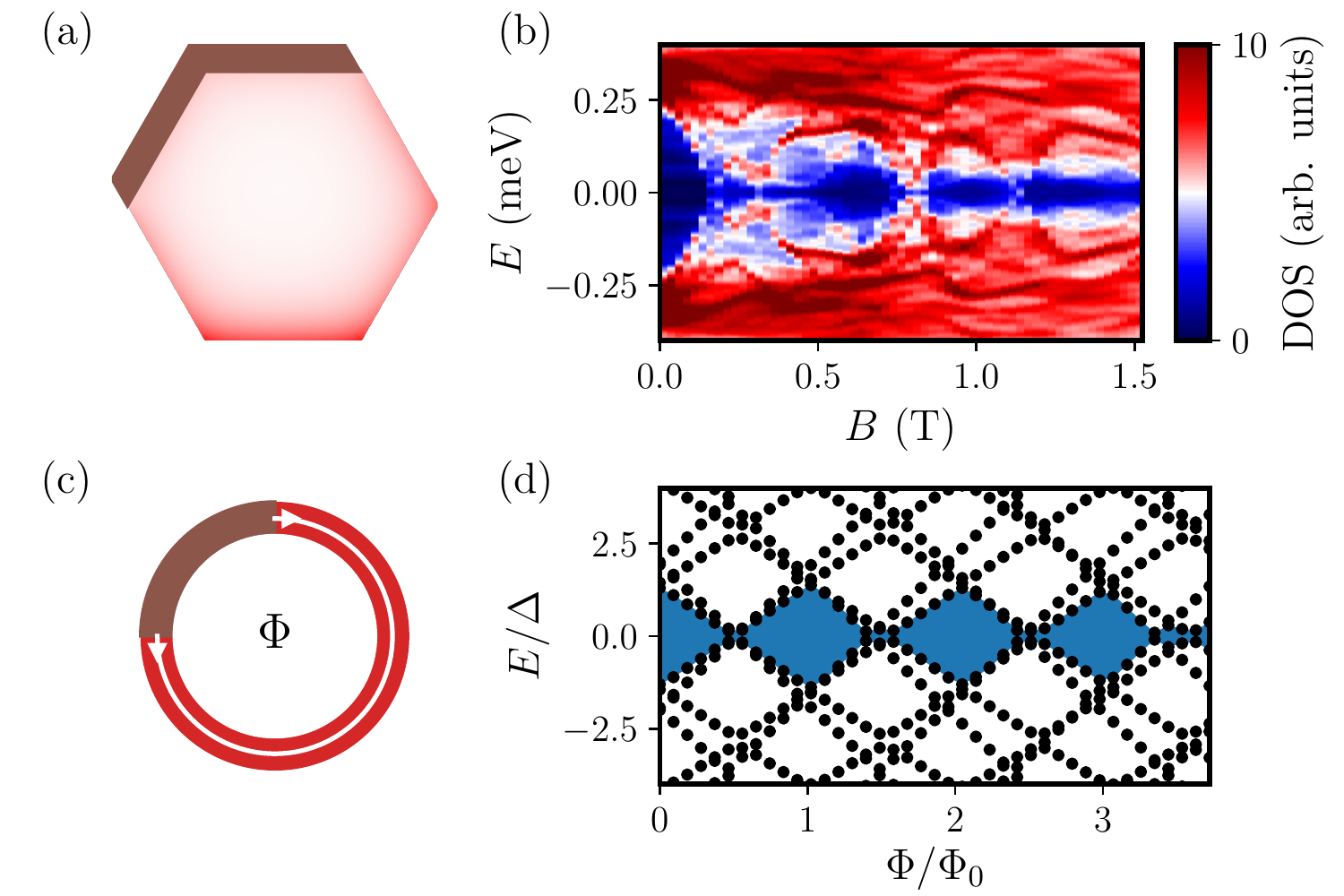}
  \caption{\label{fig:oscillation} (a) Density in two-facet device for
    $V_G=1$ V with
    $\rho_\mathrm{acc}=2\times 10^{19}$\,e/cm$^3$. (b) DOS as a
    function of magnetic field. A different colormap as in Fig.~\ref{fig:fb} is used to emphasize the oscillations. (c) Toy model geometry: red denotes semiconductor
    and brown superconductor. (d) Energy levels in toy model as a function of magnetic flux.}
\end{figure}

Because of the accumulation layer and the electrostatic screening the
system obtains a ring-like charge density for positive backgate
voltage, see Fig.~\ref{fig:oscillation}~(a) which is a special case of Fig.~\ref{fig:densities}. In this configuration
the system is very susceptible to orbital effects, resulting as we have seen in strongly enhanced $\gs$. Furthermore, we find that, in this regime, the DOS shows an
oscillatory behavior with magnetic field $B$ following the first bulk gap closing, see
Fig.~\ref{fig:oscillation}~(b). The minima correspond roughly to a
magnetic field in which half-integer magnetic flux quanta penetrate
the cross section of the wire. The maxima correspond roughly to
integer flux quanta. 

The basic mechanism of the oscillating induced gap can be understood
in a crude toy model depicted in Fig.~\ref{fig:oscillation}~(c). For
simplicity, it is assumed that the superconducting segment (shown in brown) is
longer than the coherence length. Then one can view the semiconducting
segment as a Josephson junction, in which the Andreev bound states
depend on the phase controlled by the magnetic flux penetrating the
ring~\footnote{Note that similar toy models have been investigated in Ref.~\cite{Buttiker1986, Sticlet2017}, although the
  geometry was not motivated by the electrostatics there.}. The
resulting spectrum is plotted in
Fig.~\ref{fig:oscillation}~(d). The minima correspond to values of
flux in which this phase is a half-integer multiple of $2\pi$, the
maxima to integer multiple of $2\pi$. In this toy model, the wave functions are all sensitive to the same magnetic flux, leading to very regular oscillations of the energy gap. On the other hand, in the realistic simulations different wave function can have a different effective area, and thus the oscillations are much less regular. In particular, after the first bulk gap closing in Fig. 10b, the bulk energy gap never quite recovers to the initial value. For details about the toy model
calculation see App.~\ref{sec:toy}.

\subsection{\label{sec:pd}Phase diagram}

Now we study the topological phase diagram in the presence of
electrostatic, orbital and renormalization effects. Early attempts of
calculating the topological phase diagram of Majorana nanowires
typically did not include electrostatics effects, but show the phase
boundaries as a function of the chemical potential $\mu$ rather than
the gate voltage~\cite{Lutchyn2010, Oreg2010, Tewari2012,
  Stanescu2011, Bas2016}. Only recently phase diagrams as a
function of a gate voltage have been obtained~\cite{Vuik2016, Antipov2018}. The orbital effect on the phase diagram has
been studied in Ref.~\cite{Bas2016}, although without including
electrostatic effects. In this section, we consider clean semiconductor-superconductor heterostructure, i.e. no disorder.

\begin{figure}
  \includegraphics[width =\linewidth]{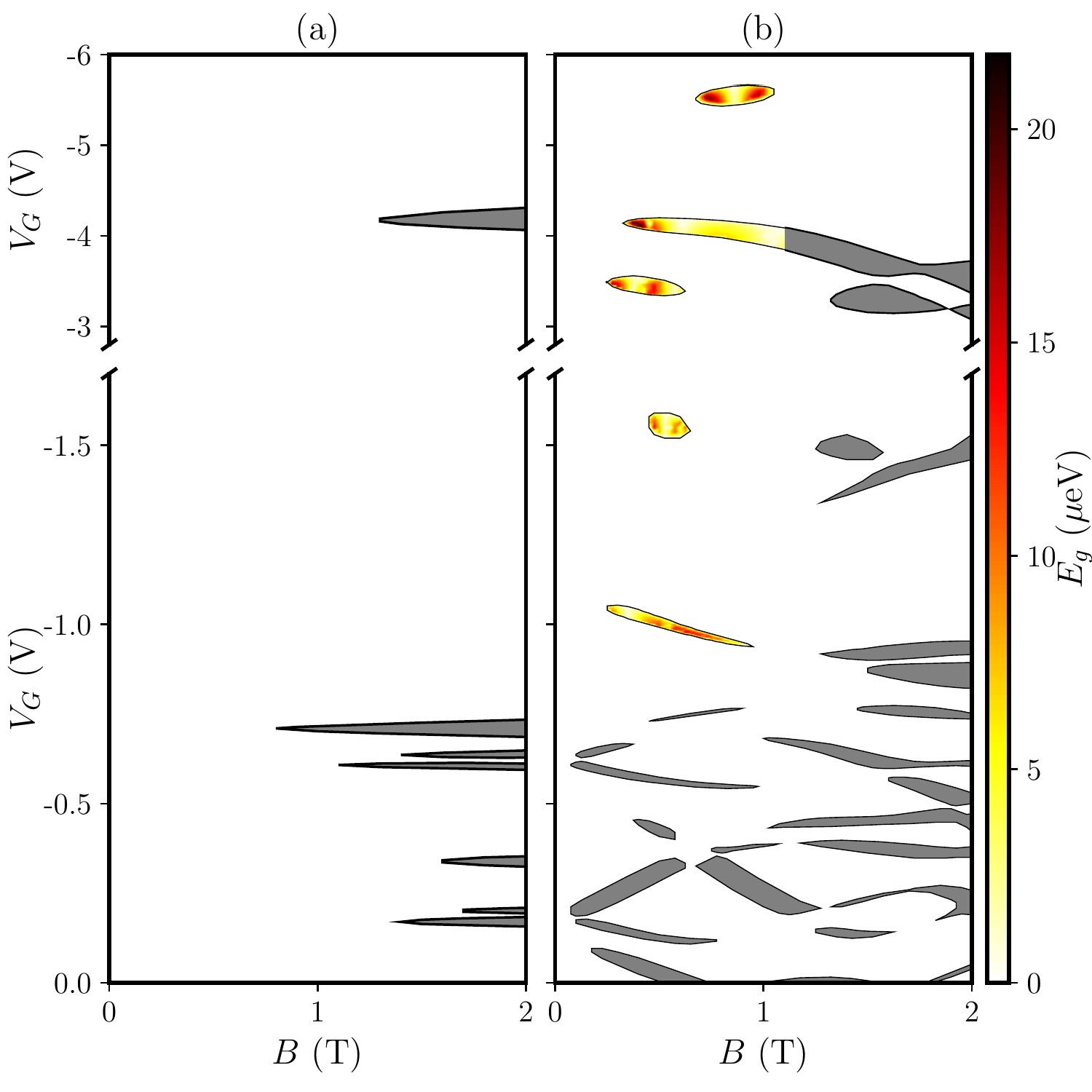}
  \caption{\label{fig:pd} Topological phase diagram of the three-facet
    device with 7 nm Al shell and
    $\rho_\mathrm{acc}=2\times 10^{19}$\,e/cm$^3$ for (a)
    without orbital effects, (b) with orbital effects. For case (b)
    the topological gap is overlaid in selected regions (it has not
    been calculated in the gray regions). 
    }
\end{figure}

In Fig.~\ref{fig:pd}~(a) we show the phase diagram of the three-facet device without orbital
effects. The phase diagram looks similar to earlier
findings of multi-subband wires~\cite{Tewari2012, Stanescu2011,
  Antipov2018}, although we find a strong dependence of the semiconductor-superconductor coupling on the subband, resulting in a large variation of minimal critical magnetic fields corresponding to the phase transition.  As has also been pointed out in
Ref.~\cite{Antipov2018}, the lever arm of chemical potential vs gate
voltage is significantly larger at positive or small negative gate
voltages than at large negative ones. Consequently, the density of
topological phases is higher in $V_G$ for small negative gate voltages
in Fig.~\ref{fig:pd}. The reason for this is twofold: first, the electron states
localized near the gate are more easily tuned by the backgate than the 
states close to the superconductor. Second, the screening effect of
the holes decreases the lever arm further for large negative gate
voltage. In general, not taking orbital effect into account leads often to
magnetic fields, at which the topological phase transitions, being large compared to experiments.

Turning the orbital effect on in Fig.~\ref{fig:pd}~(b) changes the shapes of the phase
boundaries dramatically. For small negative gate voltages the phase
diagram is dominated by the orbital effect of magnetic field. This
becomes apparent due to the small magnetic fields at which the
topological transition occur and the very non-parabolic shape of the
phase boundaries. In this regime one often finds two topological
regions emerging close in gate voltage at similar magnetic fields,
that separate from each other, one drifting to larger gate voltages
and the other to smaller gate voltages. These result from two subbands
that are near angular momentum eigenstates, with approximately
opposite angular momentum~\cite{Winkler2017}. One of the reasons why the
orbital effect is so strong is the high electron density, which is a result of the large
band-offset of InAs/Al and the accumulation layer resulting in about ten occupied subbands in InAs. High
subbands have high orbital quantum numbers coupling strongly to magnetic field~\cite{Winkler2017}. At large negative gate voltage the orbital effect is 
suppressed and the phase boundaries look closer to the ones without
orbital effect, although the influence of the orbital effect is still
strongly present.

From Fig.~\ref{fig:pd}~(b) it becomes
apparent that only topological phases with appreciable negative
backgate voltage have a sizable topological gap. We find that the
maximum topological gap is only slightly larger than
20\,$\mu$eV. While this seems like a small value we emphasize that it is proportional to the strength of the Rashba spin-orbit coupling. In our calculation, the value of spin-orbit coupling is conservatively since we take only electrostatic origin of spin-orbit coupling int account. The value of $\alpha$ we obtain from
Eq.~\eqref{eq:rashba} is typically about 10\,meV$\cdot$nm, whereas
experiments report values in the range of
10~to~30\,meV$\cdot$nm~\cite{Roulleau2010, Dong2012} which would result in a significantly larger topological gap.

\subsection{Effect of broken mirror symmetry in the two-facet device}

In terms of symmetries, the most significant difference between the
three- and two-facet devices is the vertical mirror symmetry in the
$(y,z)$-plane $\mathcal{M}_{yz}$. Additionally considering the
particle-hole symmetry $\mathcal{P}H(k)\mathcal{P}^{-1}=-H^*(-k)$,
$\mathcal{P}^2=+1$, which protects the MZMs, and the time-reversal
symmetry $\mathcal{T}H(k)\mathcal{T}^{-1}=H^*(-k)$, $\mathcal{T}^2=-1$
it can be shown that the combination of the three symmetries create a
chiral symmetry
\begin{equation}
  \mathcal{C} H(k) \mathcal{C}^{-1} = -H(k),\ \mathcal{C}^2=+1,
\end{equation}
that survives at finite magnetic field parallel to the $(y,z)$ mirror
plane. For the specific case of our Hamiltonian Eq.~\eqref{eq:ham} the
chiral symmetry is given by
$\mathcal{C}= \tau_y \sigma_z\delta(x+x')$ (with $\delta(x+x')$ being the real-space reflection operator taking $x$ to $-x$). Note that the Rashba term
$\alpha_x$ breaks this chiral symmetry.

\begin{figure}
  \includegraphics[width = \linewidth]{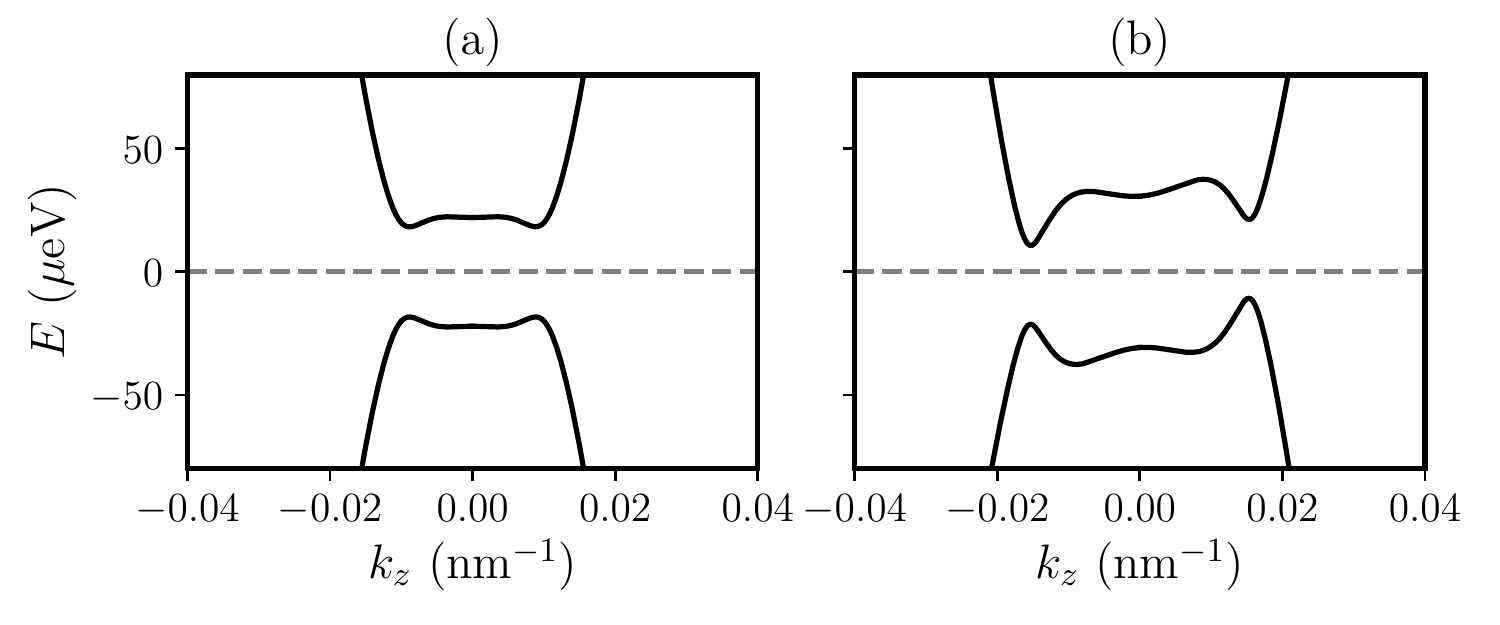}
  \caption{\label{fig:dispersion} (a) [(b)] Band structures in the
    topological phase for the three-facet [two-facet] device with the
    parameters $V_G=-4.15$\,V and $B=0.375$\,T ($V_G=-1.22$\,V and
    $B=0.575$\,T).}
\end{figure}

One particular consequence of the chiral and particle-hole symmetry is
that the band structure is line-reflection symmetric around the $k=0$
and $E=0$ axes, see Fig.~\ref{fig:dispersion}~(a). In the two-facet device the chiral symmetry is broken because of the missing
mirror symmetry $\mathcal{M}_{yz}$. Therefore, the band structure is
only point-inversion symmetric around the $(E=0, k=0)$ point, as
dictated by the particle-hole symmetry. At finite $B$ this generically leads to a
tilting of the band structure that is detrimental to the topological
gap, as is shown in Fig.~\ref{fig:dispersion}~(b). At $B=0$ the time-reversal symmetry 
prevents any tilting of the band structure, see Fig.~\ref{fig:igtf}. It has been pointed
out in Ref.~\cite{Bas2016} that a possible mirror perpendicular to the
wire axis $\mathcal{M}_{xy}$ would also prevent any tilting of the band structure. The 
wurtzite Dresselhaus term $\alpha_z$ breaks this mirror symmetry in our
case. Note that in [111] zincblende wires the Dresselhaus term 
is expected to be much smaller, due to the fact that it is
cubic in $k$ as opposed to linear in $k$ in the wurtzite case~\cite{Dresselhaus1955,Gmitra2016}. 
Therefore, it is expected that the tilting
effect of the band structure is significantly smaller in non-mirror-symmetric [111]
zincblende wires than in non-mirror-symmetric [0001] wurtzite wires, although it would still be present.

\section{Summary and Conclusions}

In this work, we presented a unified numerical approach for realistic
simulation of semiconductor-superconductor heterostructures which adequately treats all the important effects in these systems: proximity-induced superconductivity, orbital and Zeeman effect of an applied magnetic field, disorder, spin-orbit coupling as well as electrostatic environment and realistic geometry (hexagonal cross-section). We considered InAs/Al hybrid nanowires as a model system and compared our numerical results with the existing experimental data. We calculated topological phase diagrams for InAs/Al hybrid nanowires and quasiparticle gaps in corresponding topological phases which is important data for the topological qubit design proposals and can be used to optimize the designs for Majorana-based qubits. Furthermore, the approach developed here can be readily extended to other topological semiconductor-superconductor platforms such as 2DEG platforms~\cite{Shabani16, Nichele2017,
  Suominen2017}, quasi-one-dimensional networks obtained by selective area growth~\cite{SAG1,
  SAG2} or nanowires with a full superconducting shell~\cite{full_shell_experiment, full_shell_theory}.

We first investigated the dependence of the subgap density of states on the gate voltage at zero magnetic field. We identified states 
that live near the gate (i.e. opposite to the interface with Al) and within the accumulation layer of the pristine InAs surface to be responsible for a subgap density of
states for positive gate voltages. We find that at moderate negative gate voltages the dominant contribution to the subgap density of states comes from the states localized close
to the InAs/Al interface which are very strongly proximitized, resulting in a large proximity-induced gap and hard gap in the semiconductor. We also considered the effect of disorder in the superconducting shell and found that it increases the semiconductor-superconductor coupling, leading to an enhanced induced gap.

Next, we investigated the dependence of the energy spectrum on external magnetic field and studied the evolution of the spectrum as a function of electric and magnetic fields. In particular, we calculated the effective $\gs$-factor for subgap states and showed that it has a non-monotonic dependence on gate voltage, consistent with the experimental data~\cite{Sole2017}. This non-monotonic dependence appears due to the interplay of two competing effects: renormalization of the effective model parameters in the strong tunneling regime and orbital contribution to $\gs$. Therefore,
even in the strongly proximitized regime, the InAs/Al hybrid device is
able to support MZMs at moderate magnetic fields.   

We characterized topological superconductivity in hybrid InAs/Al nanowires by calculating topological phase diagram as well as the corresponding topological excitation gaps. 
Given that magnetic fields of topological phase transitions in hybrid InAs/Al nanowires are significantly smaller than expected from a simulation including only the Zeeman term, 
we show that inclusion of the orbital effect is crucial for a
quantitative and qualitative understanding of the topological superconductivity in  proximitized nanowires.

We also investigated the microscopic origin of the spin-orbit coupling in proximitized nanowires. We found that the interplay of Dresselhaus term (resulting from the wurtzite structure) and broken mirror symmetry
lead to a tilt of the band structure in the two-facet device, which is absent in mirror-symmetric three-facet devices. 
Therefore, we recommend using wires with symmetrical Al shells and gate congurations for Majorana applications, particularly in the case of wurtzite wires.

Finally, we emphasize that the level of detail of our modeling sets a new standard for simulations of hybrid semiconductor-superconductor nanowires, that should be also applied to different Majorana platforms employing semiconductor-superconductor heterostructures.
One of the remaining challenges is a fully three dimensional simulation of a hybrid device including scattering and disorder in the superconductor.
Indeed, although our two dimensional disorder model is sufficient to capture correctly some properties of the hybrid system like the magnitude of the induced gap at zero magnetic field, it is not adequate for understanding the impact of longitudinal disorder on the topological phase in the limit of strong coupling between the semiconductor and the superconductor.

\acknowledgments

We acknowledge stimulating discussions with Anton Akhmerov, Bogdan A. Bernevig, Ming-Tang Deng, Karsten Flensberg, John Gamble, Jan Gukelberger, Panagiotis Kotetes, Peter Krogstrup, Charlie Marcus, Michiel de Moor, Bas Nijholt, Enrico Rossi, Jie Shen, Saulius Vaitiek\.{e}nas and Hao Zhang. L.I.G. acknowledges the support of NSF DMR Grant No. 1603243.
 A.A.S.  acknowledges the support of
SNSF through NCCR MARVEL and QSIT grants, the SNSF Professorship and
Microsoft Research. This work was performed in part at Aspen Center for Physics, which is supported by National Science Foundation grant PHY-1607611.

\appendix

\section{\label{sec:gauge}Supercurrent distribution}

If both superconductivity and orbital effect of magnetic field are considered in the same system, it is crucial
that the combination of vector potential and superconducting phase
results in correct physical observables.  Due to the much higher density of
electrons in the superconductor compared to the semiconductor we consider the
superconductor film for these considerations to be independent of the semiconductor. In this
case, and at zero temperature, the superconductor will minimize its kinetic energy which is proportional to
$E_S \propto \int d\mathbf{r}\, J_S^2(\mathbf{r})$~\cite{Tinkham}. The correct vector potential $\mathbf{A}$ and phase can be
obtained by minimizing this expression~\cite{Wojcik2018}. Since the London penetration depth of Al is much larger than the thickness of the Al film we neglect screening effects of the supercurrent. For the simple geometry at hand we are able to write down an analytical solution for the vector potential and superconducting phase which approximates the full solution.

\begin{figure}
  \includegraphics[width = \linewidth]{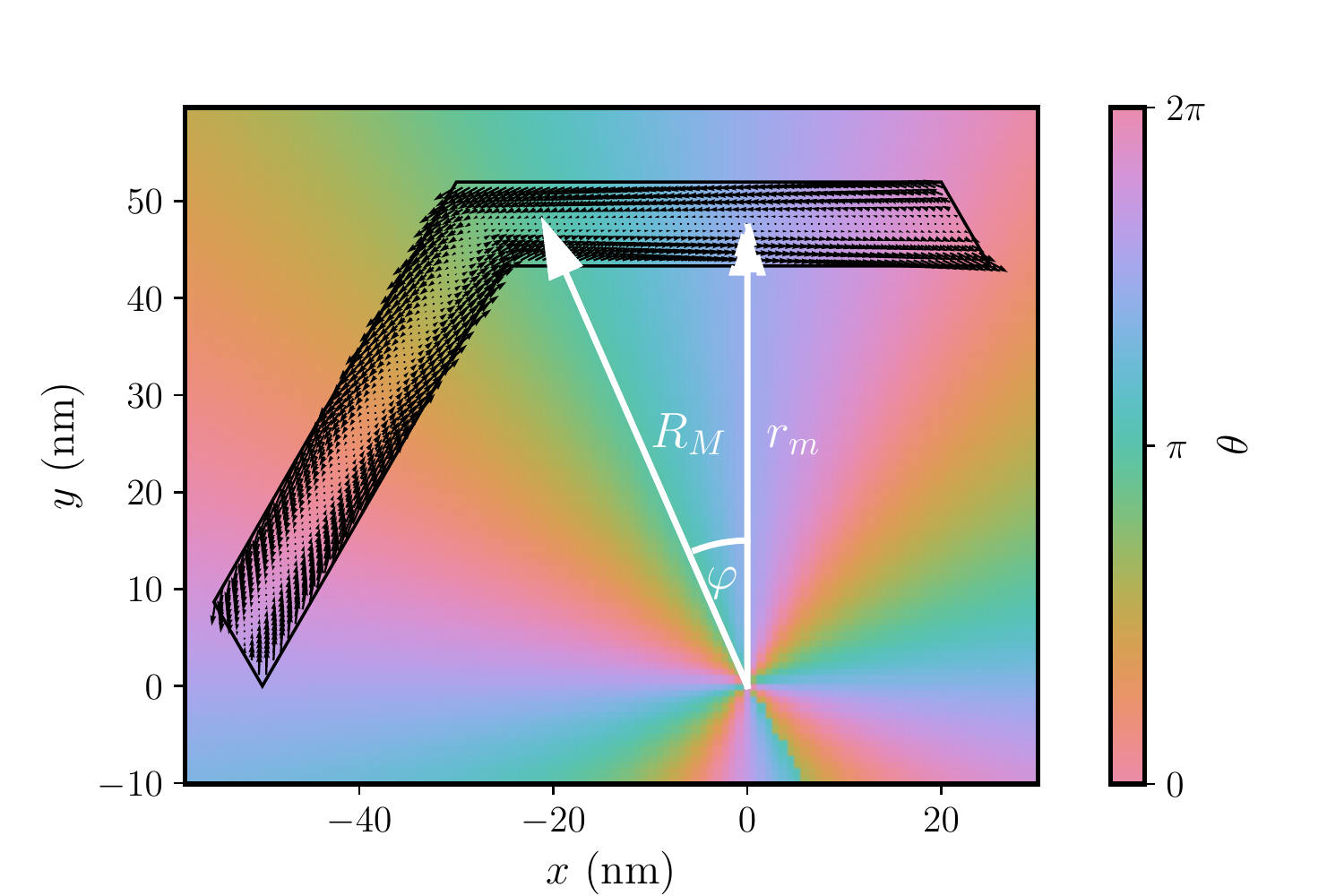}
  \caption{\label{fig:sc} Supercurrent distribution, indicated by the
    black arrows, in the superconductor at a magnetic field of
    1\,T. The background color corresponds to the phase.}
\end{figure}

We take the vector potential in the cylindrical gauge
\begin{equation}
  \mathbf{A} = \frac{B}{2} r \hat e_\varphi.
\end{equation}
Furthermore, we choose the phase $\theta$ such that the supercurrent
Eq.~\eqref{eq:sc} is canceled exactly in the middle of the
superconducting film. This is achieved by satisfying
$\frac{1}{R_M}\frac{\partial \theta}{\partial \varphi}=-\frac{2
  e}{\hbar}A_\varphi(R_M) = -\frac{2e}{\hbar}\frac{B}{2} R_M$, with
$R_M$ corresponding to the radius to the middle of the superconducting
film, see Fig.~\ref{fig:sc}. Due to the hexagonal shape of the film
$R_M$ is a function of $\varphi$: 
$R_M(\varphi) = r_m/\cos(\varphi)$ for $|\varphi| < \pi/6$ (see Fig.~\ref{fig:sc} for the definition of $r_m$). Therefore,
the phase is given by
\begin{equation}
  \theta(\varphi) = -\frac{2e}{\hbar}\frac{B}{2}\int_0^\varphi
  d\varphi' \frac{r_m^2}{\cos^2(\varphi')}= -\frac{2e}{\hbar}\frac{B}{2} r_m^2 \tan(\varphi),
\end{equation}
for $|\varphi| < \pi/6$.

In Fig.~\ref{fig:sc} we show the resulting supercurrent distribution
and phase (we continue to plot the face outside of the superconductor, even though it is strictly speaking not defined there). Since the supercurrent takes only small values the kinetic energy of the superconductor is close
to the minimum value and also the boundary conditions are fulfilled with this choice in a good approximation. Furthermore, we find that our results are
not strongly dependent on the exact choice of vector potential and phase dependence
as long as the kinetic energy is close to the minimal value. If the
magnetic flux through the hexagon of size $r_m$ is not an integer multiple of the
magnetic flux quantum the superconducting phase needs to have
a discontinuity somewhere. Since in none of our geometries all facets are
covered by the superconductor this jump can be conveniently located along a line
that does not go through a superconducting region.

\section{\label{sec:toy} Toy model}

For the toy model calculation we assume a simple BdG Hamiltonian
without spin-orbit coupling or Zeeman
\begin{equation}
  H_\mathrm{toy} = \begin{pmatrix}
    \frac{\hbar^2}{2m} k^2 -\mu & \Delta e^{i\theta} \\
    \Delta e^{-i\theta} & -\frac{\hbar^2}{2m} k^2 +\mu
  \end{pmatrix},
\end{equation}
where we use $m=0.05 m_e$, $\mu=0.5$\,eV and $\Delta=0.05$\,eV. The
vector potential is added via the substitution Eq.~\eqref{eq:ms}. For
the gauge and superconducting phase we proceed as described in
App.~\ref{sec:gauge}. For the inner radius of the ring we take 30\,nm
and for the outer radius 40\,nm. 

\section{\label{sec:thickness}Dependence of the induced gap on Al-layer thickness}

\begin{figure}
  \includegraphics[width = \linewidth]{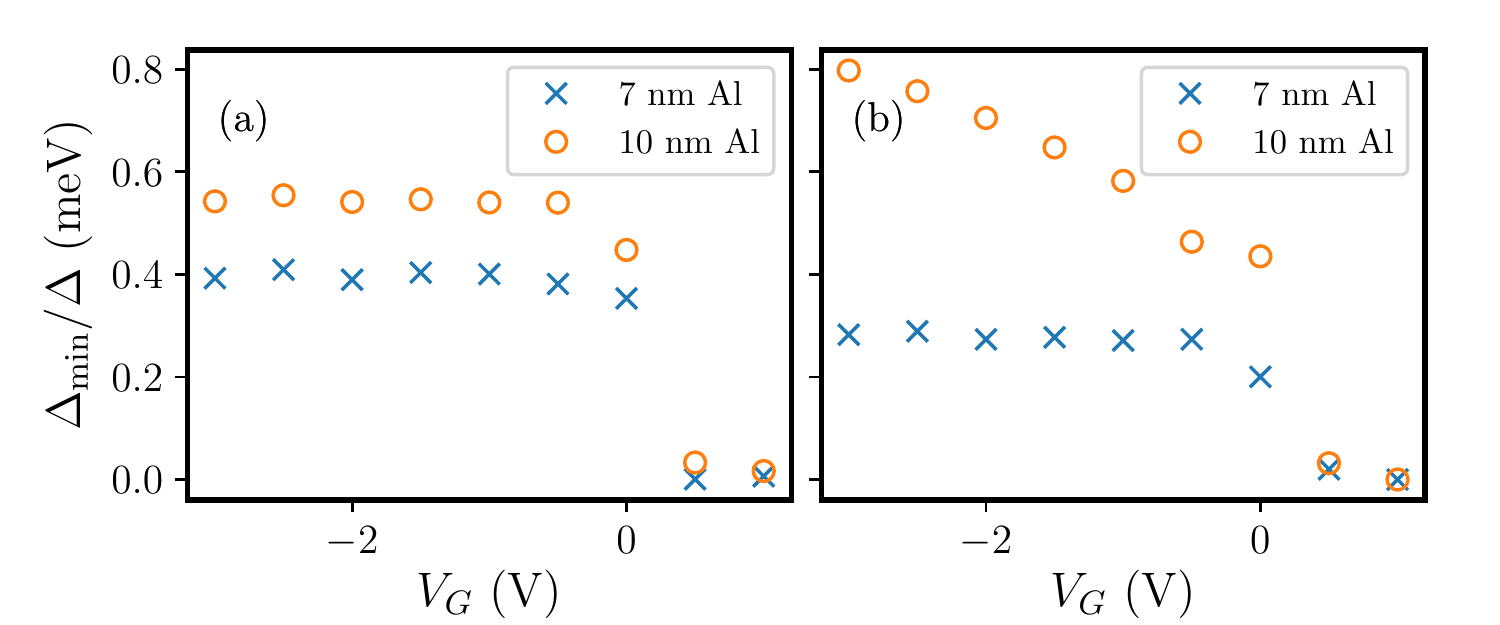}
  \caption{\label{fig:igth} Induced gap as a function of backgate
    voltage for the different devices and superconductor film thicknesses. (a)
    ((b)) Comparison of three-facet (two-facet) devices with
    $\rho_\mathrm{acc}=5\times 10^{18}$\,e/cm$^3$ for superconductor film
    thicknesses of 7 and 10 nm.}
\end{figure}

In Fig.~\ref{fig:igth} we show the induced gaps of the two-facet and three-facet device for two different thicknesses of the Al-layer: 7\,nm and 10\,nm. For the thicker Al-layer the induced gap is always larger, indicating that the semiconductor-superconductor coupling is stronger in this case. Therefore, the thickness of the superconductor-layer might be used as an additional knob, in addition to the gate~\cite{Antipov2018,Michiel2018}, to control the strength of the semiconductor-superconductor coupling.

\section{\label{sec:other} Additional phase diagrams}

Here we show additional phase diagrams calculated for other geometries and different accumulation layer strengths as in the main text. 

\begin{figure}[ht!]
  \includegraphics[width = \linewidth]{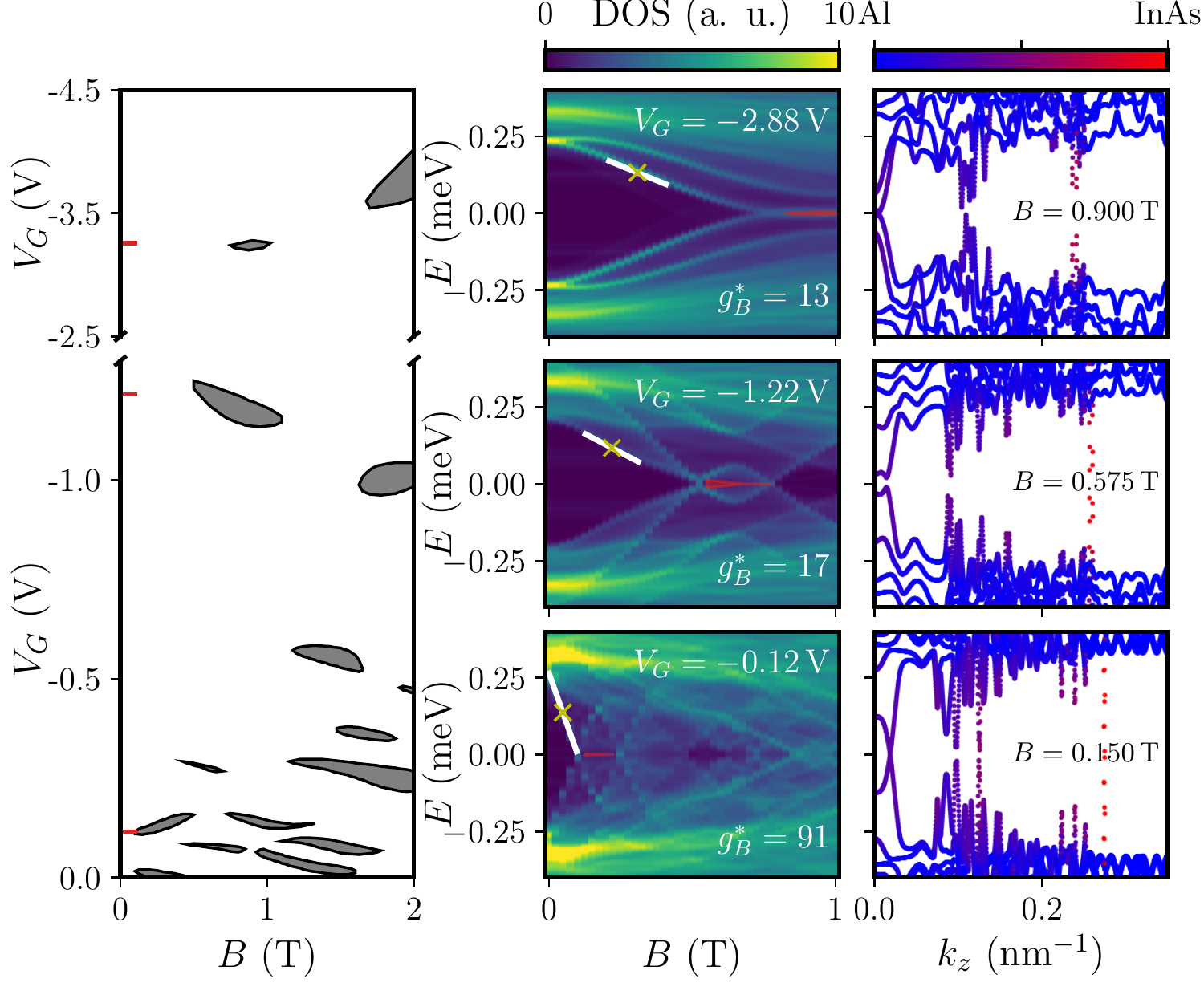}
  \caption{\label{fig:pdtf} Phase diagram, magnetic field dependent
    DOS and band structure for the two-facet device with
    $\rho_\mathrm{acc}=2\times 10^{19}$\,e/cm$^3$.}
\end{figure}

\begin{figure}[ht!]
  \includegraphics[width = \linewidth]{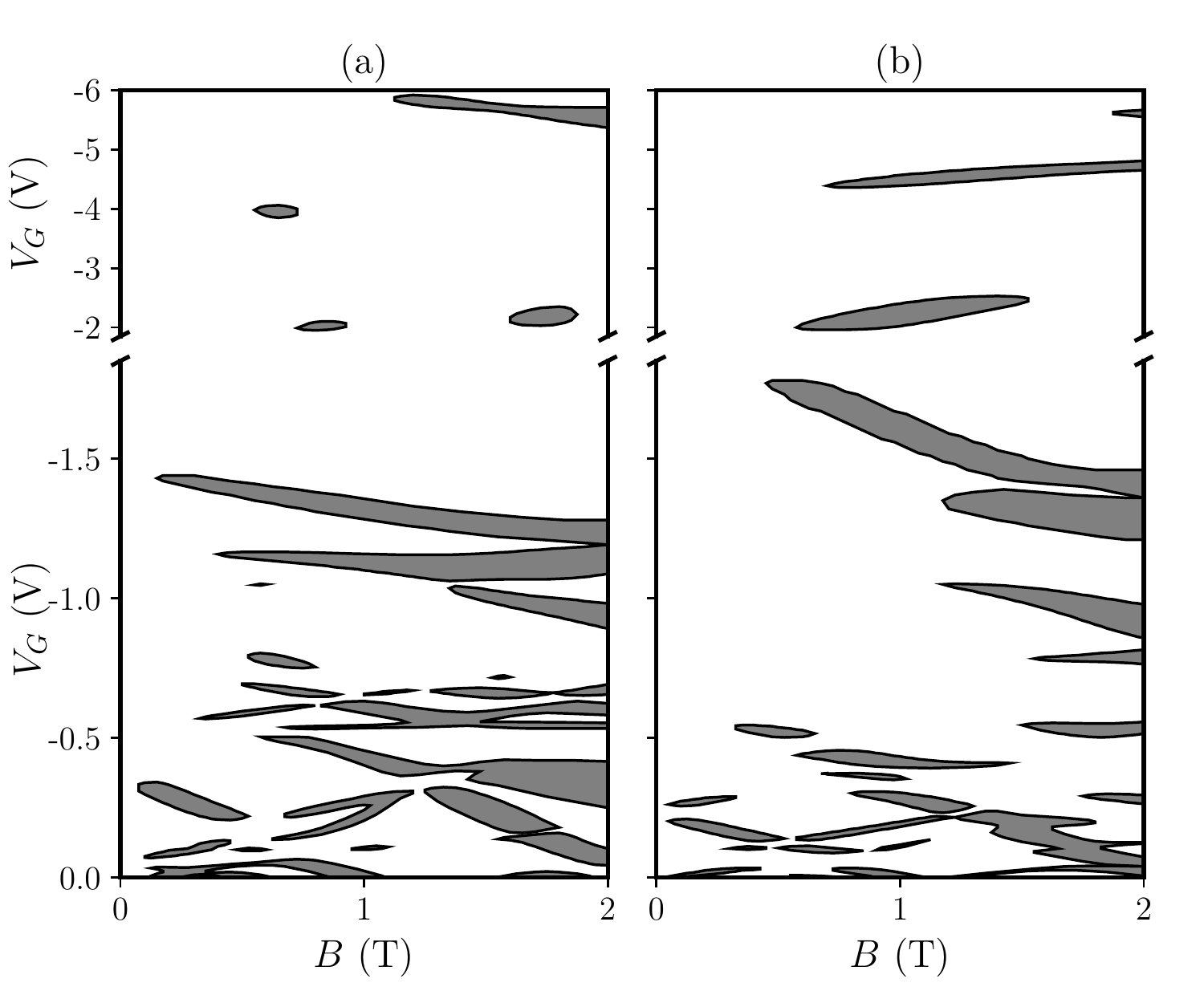}
  \caption{\label{fig:pdth} (a) ((b)) Phase diagram for the three-facet device
    with 10\,nm Al (two-facet device with 7\,nm Al) thickness and
    $\rho_\mathrm{acc}=5\times 10^{18}$\,e/cm$^3$.}
\end{figure}

In Fig.~\ref{fig:pdtf} we show the phase diagram for the two-facet wire. Since calculation of the topological gap is very expensive, we show the DOS at three cuts of different topological phases at fixed gate voltage. We find that the topological phase space is greatly reduced in comparison to the three-facet device shown in Fig.~\ref{fig:pd}~(b). Partly this is also caused by the thicker Al layer which is 10\,nm compared to 7\,nm. The effect of Al-layer thickness is further investigated in Fig.~\ref{fig:pdth}, where we show three-facet phase diagram for 10\,nm Al thickness and a two-facet phase diagram for 7\,nm Al thickness. We find that the topological phase space for the two-facet device with 7\,nm Al thickness is significantly larger than with 10\,nm Al thickness.

\begin{figure}[ht!]
  \includegraphics[width = \linewidth]{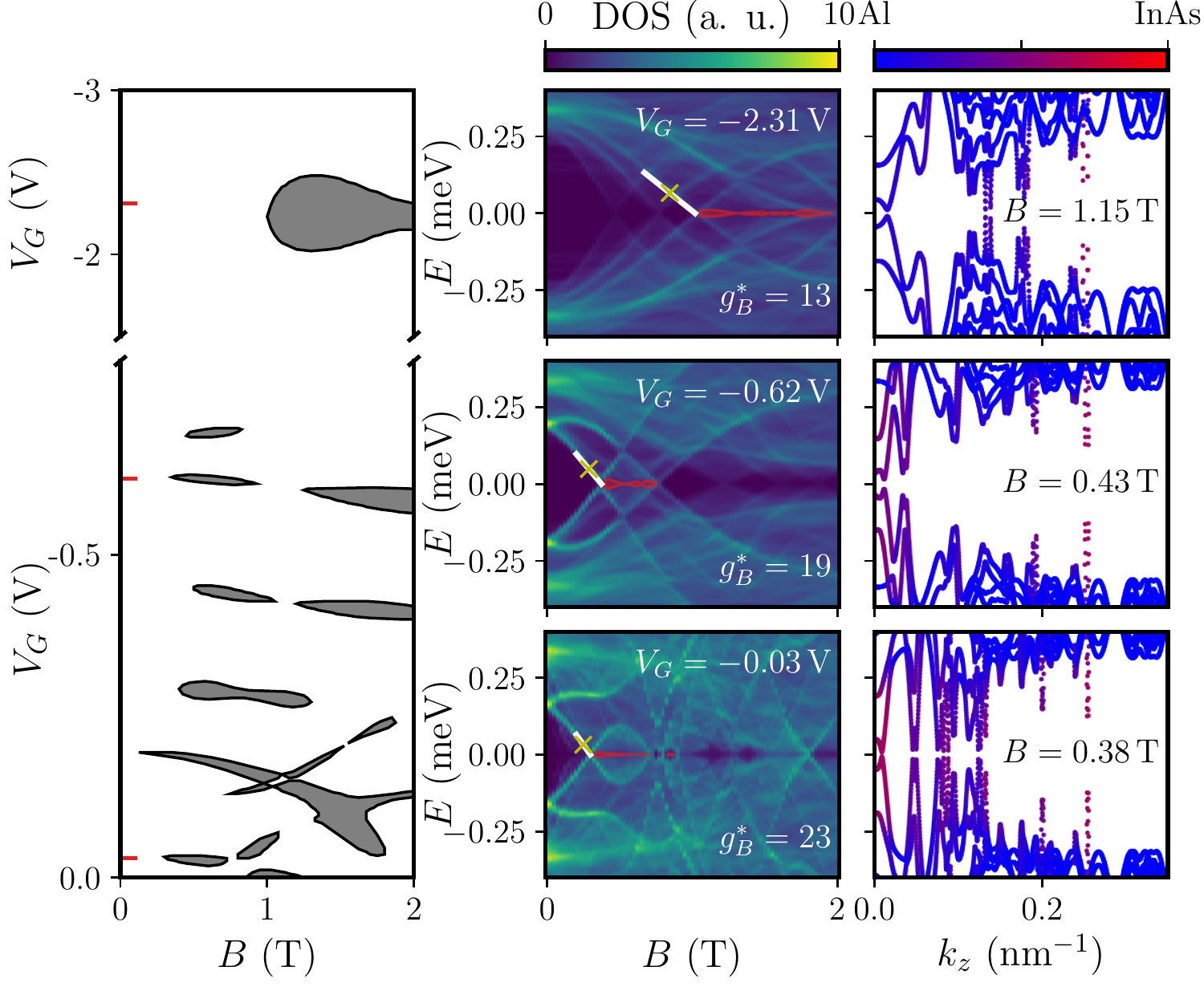}
  \caption{\label{fig:pdwa} Phase diagram, magnetic field dependent
    DOS and band structure for the three-facet device with
    $\rho_\mathrm{acc}=5\times 10^{18}$\,e/cm$^3$.}
\end{figure}

\begin{figure}[ht!]
  \includegraphics[width = \linewidth ]{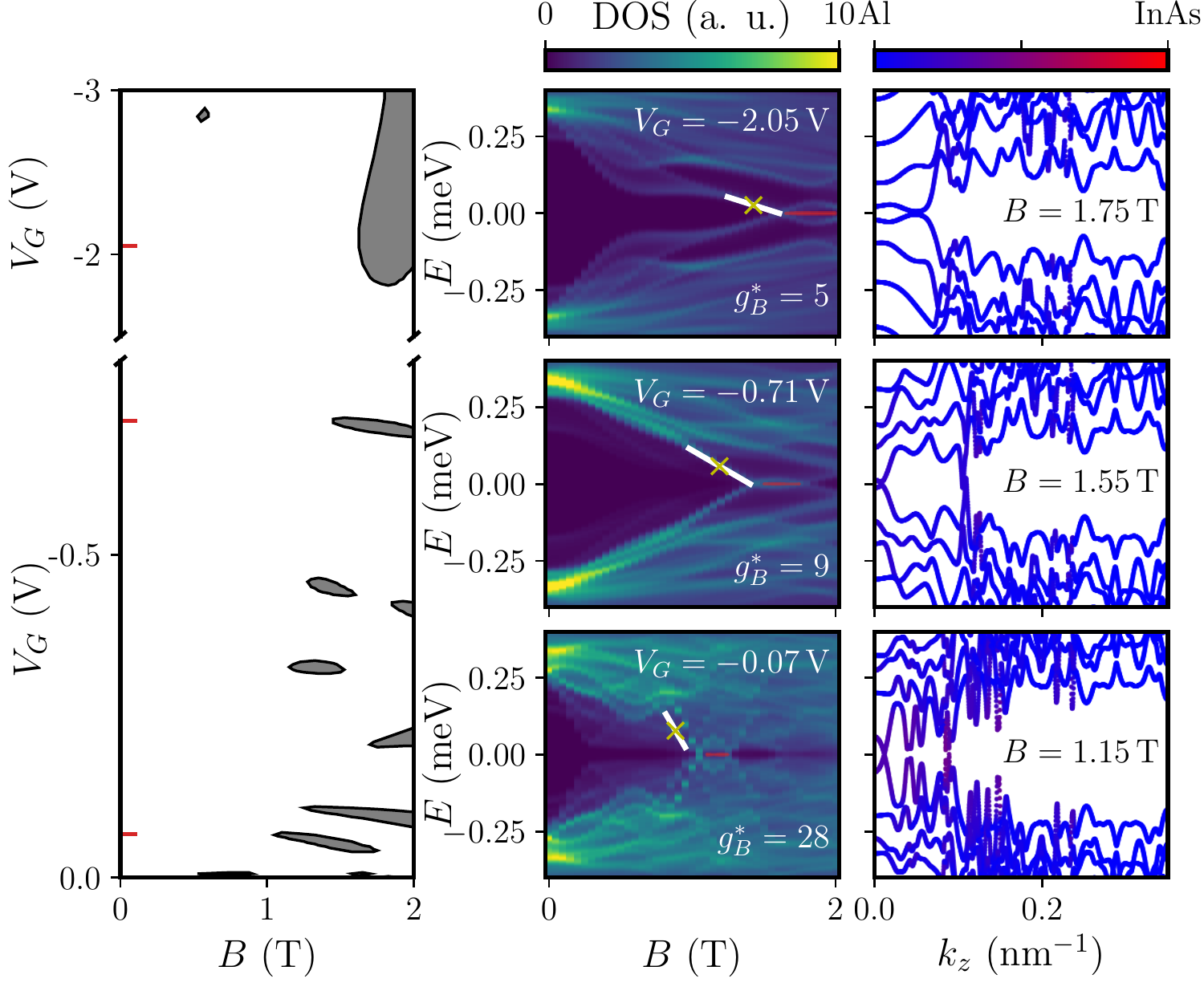}
  \caption{\label{fig:pdtfwa} Phase diagram, magnetic field dependent
    DOS and band structure for the two-facet device with
    $\rho_\mathrm{acc}=5\times 10^{18}$\,e/cm$^3$.}
\end{figure}

In Fig.~\ref{fig:pdwa} and \ref{fig:pdtfwa} we show phase diagrams for the weaker accumulation layer corresponding to $\rho_\mathrm{acc}=5\times 10^{18}$\,e/cm$^3$. 

\FloatBarrier

\bibliography{literature}

\end{document}